\makeatletter
\declare@file@substitution{revtex4-1.cls}{revtex4-2.cls}
\makeatother
\documentclass[twocolumn]{aastex631}

\usepackage{array,multirow}

\shorttitle{Relative velocities between $^{13}$CO structures in molecular clouds}
\shortauthors{Yuan et al.}
\graphicspath{{./}{figures/}}

\begin{document}

\title{Relative velocities between $^{13}$CO structures within $^{12}$CO Molecular Clouds}

\correspondingauthor{Ji Yang}
\email{jiyang@pmo.ac.cn}

\author[0000-0003-0804-9055]{Lixia Yuan}
\affiliation{Purple Mountain Observatory and Key Laboratory of Radio Astronomy, Chinese Academy of Sciences, \\
10 Yuanhua Road, Qixia District, Nanjing 210033, PR China}
\email{lxyuan@pmo.ac.cn}

\author[0000-0001-7768-7320]{Ji Yang}
\affiliation{Purple Mountain Observatory and Key Laboratory of Radio Astronomy, Chinese Academy of Sciences, \\
10 Yuanhua Road, Qixia District, Nanjing 210033, PR China}

\author[0000-0003-3151-8964]{Xuepeng Chen}
\affiliation{Purple Mountain Observatory and Key Laboratory of Radio Astronomy, Chinese Academy of Sciences, \\
10 Yuanhua Road, Qixia District, Nanjing 210033, PR China}

\author[0000-0002-0197-470X]{Yang Su}
\affiliation{Purple Mountain Observatory and Key Laboratory of Radio Astronomy, Chinese Academy of Sciences, \\
10 Yuanhua Road, Qixia District, Nanjing 210033, PR China}

\author[0000-0003-2549-7247]{Shaobo Zhang}
\affiliation{Purple Mountain Observatory and Key Laboratory of Radio Astronomy, Chinese Academy of Sciences, \\
10 Yuanhua Road, Qixia District, Nanjing 210033, PR China}

\author[0000-0003-2418-3350]{Xin Zhou}
\affiliation{Purple Mountain Observatory and Key Laboratory of Radio Astronomy, Chinese Academy of Sciences, \\
10 Yuanhua Road, Qixia District, Nanjing 210033, PR China} 

\author[0000-0003-0849-0692]{Zhiwei Chen}
\affiliation{Purple Mountain Observatory and Key Laboratory of Radio Astronomy, Chinese Academy of Sciences, \\
10 Yuanhua Road, Qixia District, Nanjing 210033, PR China}

\author[0000-0003-4586-7751]{Qing-Zeng Yan}
\affiliation{Purple Mountain Observatory and Key Laboratory of Radio Astronomy, Chinese Academy of Sciences, \\
10 Yuanhua Road, Qixia District, Nanjing 210033, PR China}

\author[0000-0001-8060-1321]{Min Fang}
\affiliation{Purple Mountain Observatory and Key Laboratory of Radio Astronomy, Chinese Academy of Sciences, \\
10 Yuanhua Road, Qixia District, Nanjing 210033, PR China}

\author[0000-0002-7489-0179]{Fujun Du}
\affiliation{Purple Mountain Observatory and Key Laboratory of Radio Astronomy, Chinese Academy of Sciences, \\
10 Yuanhua Road, Qixia District, Nanjing 210033, PR China}
 
\author[0000-0002-3904-1622]{Yan Sun}
\affiliation{Purple Mountain Observatory and Key Laboratory of Radio Astronomy, Chinese Academy of Sciences, \\
10 Yuanhua Road, Qixia District, Nanjing 210033, PR China}

\author[0000-0003-0746-7968]{Hongchi Wang}
\affiliation{Purple Mountain Observatory and Key Laboratory of Radio Astronomy, Chinese Academy of Sciences, \\
10 Yuanhua Road, Qixia District, Nanjing 210033, PR China}

\author[0000-0001-5602-3306]{Ye Xu}
\affiliation{Purple Mountain Observatory and Key Laboratory of Radio Astronomy, Chinese Academy of Sciences, \\
10 Yuanhua Road, Qixia District, Nanjing 210033, PR China}

\begin{abstract}

Velocity fields of molecular clouds (MCs) can provide crucial information on 
the merger and split between clouds, as well as their internal kinematics and maintenance, 
energy injection and redistribution, even star formation within clouds. 
Using the CO spectral lines data from the Milky Way Imaging Scroll Painting (MWISP) survey, 
we measure the relative velocities along the line of sight ($\Delta$V$_{\rm LOS}$) 
between $^{13}$CO structures within $^{12}$CO MCs. 
Emphasizing MCs with double and triple $^{13}$CO structures, 
we find that approximately 70$\%$ of $\Delta$V$_{\rm LOS}$ values are less than $\sim$ 1 km s$^{-1}$, 
and roughly 10$\%$ of values exceed 2 km s$^{-1}$, with a maximum of $\sim$ 5 km s$^{-1}$.  
Additionally, we compare $\Delta$V$_{\rm LOS}$ with the internal velocity dispersion of $^{13}$CO structures ($\sigma_{\rm ^{13}CO,in}$) 
and find that about 40$\%$ of samples in either double or triple regime display distinct velocity discontinuities,  
i.e. the relative velocities between $^{13}$CO structures are larger than the internal linewidths of $^{13}$CO structures. 
Among these 40$\%$ samples in the triple regime, 33$\%$ exhibit signatures of combinations 
through the two-body motion, whereas the remaining 7$\%$ show features of configurations through the multiple-body motion. 
The $\Delta$V$_{\rm LOS}$ distributions for MCs with double and triple $^{13}$CO structures are similar, 
as well as their $\Delta$V$_{\rm LOS}$/$\sigma_{\rm ^{13}CO,in}$ distributions.
This suggests that relative motions of $^{13}$CO structures within MCs are random and independent of cloud complexities and scales. 

\end{abstract}


\section{Introduction}\label{sec:intro}
Molecular clouds are not isolated systems, 
but rather part of a network spanning from the warm diffuse ISM to the cold and dense molecular clouds. 
The dynamic evolution of these clouds can be characterized by two aspects: 
(1) Internally, hierarchical structures form within clouds, 
potentially leading to the formation of stars in gravitationally bound substructures; 
until, eventually, stellar feedback disperses the clouds \citep{Larson1981, Myers1983, McKee1989, Ballesteros1999, MacLow2004, Molinari2010, Andre2010, Andre2014, Yuan2020}. 
(2) Externally, mergers or splits between MCs and flows of gas in and out of the atomic phase can 
transfer mass, energy, and momentum, altering the properties of clouds 
\citep{Tasker2009, Dobbs2011, Dobbs2013, Dobbs2015,Fukui2016, Gong2017, Fukui2018, Fukui2021, Jeffreson2021, Skarbinski2023, Jeffreson2024}. 
Molecular clouds thus undergo a range of dynamical processes, from the large-scale differential rotation and shearing motions 
to the sub-cloud dynamics associated with star formation and feedback \citep{Ballesteros2020, Henshaw2020}. 
The possible driving forces, such as the thermal instability \citep{Field1965, Koyama2000}, 
converging flows \citep{Vazquez1995, Passot1995, Parades1999, Heitsch2006, Beuther2020}, 
gravity \citep{Lin1964, Goldreich1965, Vazquez2007, Traficante2020}, 
and magnetic fields \citep{Chandrasekhar1953, Shu1987, Myers1988a, Semadeni2011, Li2018} 
likely play a role at various stages of these processes.  
However, the details of the MC evolution, both internally and externally, remain incompletely understood. 

The properties of MCs have been extensively researched and analyzed in the literature. 
This includes their masses, sizes, velocity dispersion, etc., 
and the scaling relations between cloud sizes, velocity dispersion, surface densities, etc. 
\citep[e.g.][]{Larson1981, Dame1986, Solomon1987, Heyer2009,Roman-Duval2010,Rice2016, Miville2017, Riener2020, Rani2023}. 
Moreover, the column density probability distribution \citep[e.g.][]{Vazquez-Semadeni1994, Ma2021, Ma2022} and the velocity structure function 
\citep[e.g.][]{Heyer2004, Heyer2006} are also used to investigate the distributions of column densities and velocity fields in MCs. 
These studies provide a comprehensive understanding of the kinematic, dynamic, structural, and evolutionary characteristics of MCs. 

Our recent series of studies based on a large sample of MCs from the Milky Way Imaging 
Scroll Painting (MWSIP) CO survey \citep{Su2019}, find that as MCs grow in scale, 
they tend to exhibit more complex filamentary networks \citep[hereafter Paper I]{Yuan2021}
and host greater numbers of $^{13}$CO structures \citep[hereafter Paper II]{Yuan2022}. 
These $^{13}$CO structures have areas that generally do not exceed 70$\%$ of the MC's $^{12}$CO emission areas (Paper II). 
In addition, we revealed a preferred spatial separation among individual $^{13}$CO structures, which is independent of the MC's scale \citep[hereafter Paper III]{Yuan2023}. 
Furthermore, we found that the relative motions between $^{13}$CO structures are the primary contributor to the total velocity dispersions of MCs \citep[hereafter Paper IV]{Yuan2023b}.
Based on these observed results, an alternative picture for the assembly and destruction of MCs has been proposed. 
It suggests that the regularly spaced $^{13}$CO structures serve as the building blocks for MCs, 
and the transient processes of MCs occur through slow mergers or splits among these fundamental blocks. 
Meanwhile, these processes do not significantly alter MCs' density structures but do affect their global velocity fields. 
Numerical models indicate that mergers between clouds are gentle and do not heavily impact the density structures of clouds, 
but they can result in higher velocity dispersions of MCs \citep{Dobbs2011, Dobbs2015, Jeffreson2021, Skarbinski2023}, 
which are consistent with our pictures. 

The picture above provides crucial insights into the build-up process of MC. 
It is mainly based on the spatial distribution and motions of internal $^{13}$CO structures. 
The motions of gas displays distinct characteristics, such as noticeable variations in velocity, 
which likely result from the process of merging or splitting.  
$^{13}$CO structures are considered as the fundamental building blocks of material transfer between clouds as described in Paper III and Paper IV. 
By analyzing velocity fields of $^{13}$CO structures within clouds, 
we can gain key clues to resolve the gas motions within clouds and better understand 
the material transfer processes between clouds. 
It also sheds light on the energy injection sources driving the kinematics of MCs, 
and the processes of energy redistribution and material gathering for star formation. 

In this research, we focus on the relative velocities of the $^{13}$CO structures. 
This paper is organized as follows: 
Section 2 presents the data from the MWISP CO survey, along with the identification of $^{12}$CO molecular clouds and 
their harbored $^{13}$CO structures. 
Section 3 mainly describes the results, including the distributions of line-of-sight relative velocities between $^{13}$CO structures within MCs, 
as well as ratios between relative velocities and velocity dispersions of $^{13}$CO structures within MCs, 
the fractions of MCs displaying distinct velocity discontinuities. 
In Section 4, we discuss the observational bias in our results and  
compare our observed results with previously simulated works. 
Section 5 summarizes our findings.

\section{Data}
\subsection{The $^{12}$CO(J=1-0) and $^{13}$CO(J=1-0) spectral lines data from the MWISP survey} 
The Milky Way Imaging Scroll Painting (MWISP) survey is an ongoing northern Galactic plane CO survey, 
which is performed by the 13.7m telescope at Delingha, China, and observes the $^{12}$CO, $^{13}$CO, and C$^{18}$O 
lines at the transition $J$=1-0, simultaneously.
A detailed description of the performance of the telescope and its 3$\times$3 multibeam sideband-separating 
Superconducting SpectroScopic Array Receiver (SSAR) system is given in \cite{Su2019, Shan2012}.  
The observational strategy and raw data processing are also introduced in \cite{Su2019}. 
The half-power beamwidth (HPBW) of the antenna at the frequencies of 115 GHz is $\sim$ 50$^{\prime \prime}$. 
The typical system temperature is $\sim$ 250 K at a line frequency of the $^{12}$CO line (115.271 GHz) in the upper sideband and 
$\sim$ 140 K at $^{13}$CO (110.201 GHz) and C$^{18}$O (109.782 GHz) lines in the lower sideband, respectively. 
The total bandwidth of 1 GHz with 16,384 channels provides a spectral resolution of 61 kHz per channel, 
resulting in a velocity resolution of about 0.16 km s$^{-1}$ for $^{12}$CO lines and 0.17 km s$^{-1}$ for $^{13}$CO and C$^{18}$O lines.  
The typical RMS achieved in $^{12}$CO and $^{13}$CO lines are $\sim$ 0.5 K and $\sim$ 0.3 K, respectively. 

In this work, the $^{12}$CO and $^{13}$CO lines data are from the MWISP survey and cover about 450 deg$^{2}$ region 
with the Galactical longitude $l$ from 104$^{\circ}$.75 to 150$^{\circ}$.25, the Galactical latitude $|b| < 5^{\circ}.25$, 
and the line-of-sight velocity of $-$95 km s$^{-1}$  $<$ V$_{\rm LSR}$ $<$ 25 km s$^{-1}$.  
These $^{12}$CO and $^{13}$CO lines emission data also have been analyzed in our previous series of works in \cite{Yuan2021, Yuan2022, Yuan2023, Yuan2023b}. 
 
\subsection{The $^{12}$CO molecular clouds and their internal $^{13}$CO structures}
In our analysis, the $^{12}$CO molecular cloud is defined as a set of adjacent voxels in the position-position-velocity (PPV) space 
with observed $^{12}$CO(1-0) line intensities exceeding a certain threshold. 
The Density-based Spatial Clustering of Applications with Noise (DBSCAN) algorithm, 
designed to discover clusters with arbitrary shapes in large spatial databases \citep{ester1996}, 
was employed to identify MCs in the $^{12}$CO data cube by \cite{Yan2020}.
This algorithm combines both intensity levels and continuity of signals, 
which is appropriate for the extended and irregular shapes of MCs. 
Three parameters are used in the DBSCAN to extract the $^{12}$CO MCs, 
one parameter of \textit{cutoff} determines the line intensity threshold, 
while the other two parameters of $\epsilon$ and \textit{MinPts} define the connectivity of the extracted structures. 
A core point within extracted structures satisfies that its adjacent points within a certain radius ($\epsilon$) have 
to exceed a threshold number (\textit{MinPts}). 
The border point is inside the $\epsilon$-radius of a core point, 
but its adjacent points within the radius of $\epsilon$ do not exceed the number of \textit{MinPts} \citep{Yan2020}.  
The parameters of \textit{cutoff}=2$\sigma$ ($\sigma$ is the rms noise, whose value is $\sim$ 0.5 K for the 
$^{12}$CO line emission), \textit{MinPts}=4, and $\epsilon$=1 are used for the identification of $^{12}$CO 
clouds, as suggested in \citep{Yan2020}.  
Moreover, post-selection criteria are utilized to avoid noise contamination, 
which includes: (1) the total number of voxels in each extracted structure is greater than 16; 
(2) the peak intensity of extracted voxels must be higher than the \textit{cutoff} value adding 3$\sigma$;
(3) the angular area of the extracted structure must be larger than one beam size (2$\times$2 pixels $\sim$ 1$^{\prime}$);
and (4) the number of velocity channels must be greater than 3. 
Using aforementioned parameters and criteria, a catalog of 18,190 $^{12}$CO molecular clouds was identified in the above region by the DBSCAN algorithm \citep{Yan2021}.  
We have visually inspected and classified these 18,190 MCs into filaments and nonfilaments in Paper I. 
Additionally, the dependence of extracted MC samples on the finite angular resolution, sensitivity of observed 
spectral lines, and different algorithms have been systematically investigated in \cite{Yan2022}.

Individual $^{13}$CO structures are defined as connected voxels in the PPV space, meanwhile the $^{13}$CO 
line intensities on these voxels must exceed a certain threshold. 
We have extracted the $^{13}$CO structures within the boundaries of $^{12}$CO MCs using the DBSCAN algorithm. 
The utilized parameters are same with the parameters for $^{12}$CO MC identification, 
with the exception of the post-selection criteria of the peak intensities exceeding the \textit{cutoff} value adding 2$\sigma$, 
where $\sigma$ is $\sim$ 0.25 K for $^{13}$CO spectral lines.
Among a total of 18,190 $^{12}$CO clouds, 2851 $^{12}$CO clouds are identified to harbor the $^{13}$CO structures (Paper II). 
We have presented the distributions of the physical properties of these extracted $^{13}$CO structures in Paper II and III.
Among our 2851 MCs having $^{13}$CO structures, 
1848 (64.8$\%$) MCs have a single $^{13}$CO structure, 443 (15.5$\%$) MCs have double $^{13}$CO structures, 
185 (6.5$\%$) MCs have triple $^{13}$CO structures, the rest (13.2$\%$) have more than three $^{13}$CO structures, 
as listed in Table \ref{tab:t_cloud}.

In addition, according to the spiral structure model of the Milky Way \citep{Reid2016, Reid2019, Xu2023}, the MC samples are further divided into two groups, 
i.e., near and far. The MC samples in the near group have central velocities ranging from -30 to 25 km s$^{-1}$, 
which are mainly distributed in the local arm and have kinematical distances of $\sim$ 0.5 kpc. 
The MC samples in the far group have central velocities in a range of (-95 -30) km s$^{-1}$, 
most of which are located in the Perseus arm and their kinematical distances concentrate on $\sim$ 2 kpc \citep{Reid2016}. 
Considering these kinematical distances, 
the physical scale for the MC with an angular size of 1$^{\prime}$ in the local arm is about 0.15 pc, 
and this value is $\sim$ 0.6 pc for that in the perseus arm. 
The number of MC samples distributed in the near and far groups has also been listed in Table \ref{tab:t_cloud}.

\begin{deluxetable}{lcccc}
    \tablecaption{Number of MC sources in different groups. \label{tab:t_cloud}}
    \tablewidth{0pt}
    \tablehead{\colhead{Group} & \colhead{Sources} & \colhead{Sources in Near} & \colhead{Sources in Far} \\}
    \startdata
    All & 2851 & 988 & 1863    \\
    Single & 1848 & 576 & 1272 \\
    Double & 443 & 163 & 280 \\
    Triple & 185 & 80 & 105 \\
    \enddata
\end{deluxetable}

\section{Relative velocities between $^{13}$CO structures within molecular clouds}

In our previous studies (Paper III and IV), 
we have proposed that regularly spaced $^{13}$CO structures serve as the fundamental units of the gas transfer 
between clouds. Additionally, mergers or splits between clouds have an impact on their velocity fields, resulting in two distinct features: 
(1) relatively discontinuous velocity fields with blue and redshifted velocities, and 
(2) velocity structures with two or more velocity components, leading to higher velocity dispersions. 
Based on these features, we aim to investigate the relative velocities between individual $^{13}$CO structures within 
MCs and the imprints on the build-up processes of MCs they provide. 

To achieve this, we first determine the relative velocities between each pair of $^{13}$CO structures within the MCs. 
The relative velocities between $^{13}$CO structures are defined as the absolute differences between 
their centroid velocities.  
Thus the relative velocity ($\Delta$V$_{\rm LOS}$) between $j$th and $(j-1)$th $^{13}$CO structures along the line of sight can be calculated as: 
\begin{eqnarray} \label{e:vdiff}
\Delta V_{j, j-1} &=& \left| V_{\rm cen,^{13}CO,j} - V_{\rm cen,^{13}CO,j-1} \right| \nonumber \\
V_{\rm cen,^{13}CO,j} &=& \Sigma^{j\rm th}_{i} T_{\rm ^{13}CO,ji} V_{\rm ^{13}CO,ji}/\Sigma^{j\rm th}_{i} T_{\rm ^{13}CO,ji}, \nonumber \\
\end{eqnarray}
where V$_{\rm cen,^{13}CO,j}$ represents the centroid velocity of the $j$th $^{13}$CO structure, 
the sum $\Sigma^{j\rm th}$ runs over all voxels within the $j$th $^{13}$CO structure, 
the T$_{\rm ^{13}CO,ji}$ and V$_{\rm ^{13}CO,ji}$ are the brightness temperature and line-of-sight velocity of 
$^{13}$CO emission at the $i$th voxel in the $j$th $^{13}$CO structure.

Secondly, we compare the relative velocities with the internal velocity dispersion of $^{13}$CO structures. 
The internal velocity dispersion of $^{13}$CO structures ($\sigma_{\rm ^{13}CO,in}$) within a cloud are defined as:
\begin{eqnarray} \label{e:vdisp}
\sigma_{\rm ^{13}CO,in}^{2} & = & \Sigma^{\rm cloud}_{j} F_{\rm ^{13}CO,j} \sigma_{\rm ^{13}CO,j}^{2} /\Sigma^{\rm cloud}_{j} F_{\rm ^{13}CO,j}, \nonumber \\
\sigma_{\rm ^{13}CO,j}^{2} & = & \Sigma^{j\rm th}_{i} T_{\rm ^{13}CO,ji}(V_{\rm ^{13}CO,ji}-V_{\rm cen,^{13}CO, j})^{2}/\Sigma^{j\rm th}_{i} T_{\rm ^{13}CO,ji}, \nonumber \\
\end{eqnarray}
where the sum $\Sigma^{\rm cloud}_{j}$ runs over the whole individual $^{13}$CO structures within a $^{12}$CO cloud, 
the $\sigma_{\rm ^{13}CO,j}$ is the velocity dispersion within the $j$th $^{13}$CO structure, 
$F_{\rm ^{13}CO, j}=\int T_{mb}(l,b,v)dldbdv=0.167 \times 0.25 \Sigma_{i}^{jth} T_{\rm ^{13}CO,ji}$ (K km s$^{-1}$ arcmin$^{2}$) 
is the integrated flux of $^{13}$CO line emission for the $j$th $^{13}$CO structure. 

Furthermore, if the relative velocities between $^{13}$CO structures within a cloud satisfy: 
\begin{equation}
    \Delta V_{\rm LOS} > \sqrt{8\rm ln2}\sigma_{\rm ^{13}CO,in}
\end{equation}
i.e. the relative velocities between $^{13}$CO structures are greater than the internal linewidth of 
$^{13}$CO structures, they are characterized as the distinct velocity discontinuities ($\Delta V_{\rm dis}$). 

\begin{figure*}[ht]
    \plotone{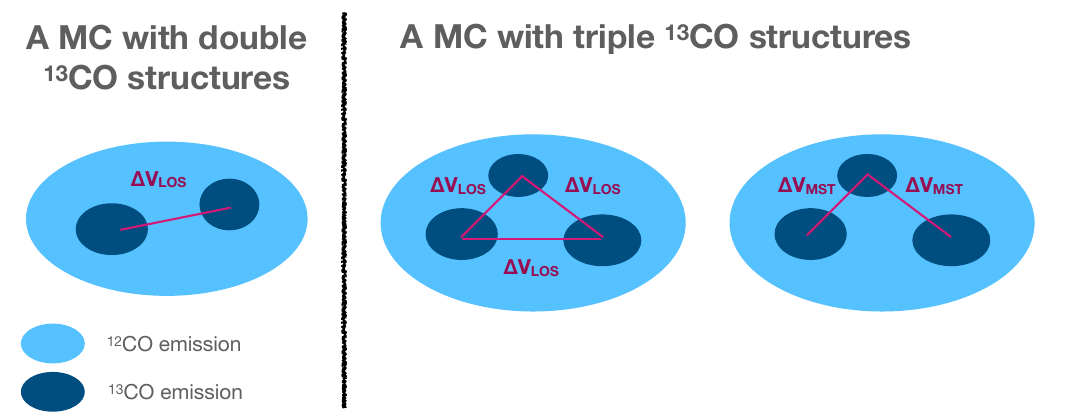} 
    \caption{Relative velocities along the line of sight between $^{13}$CO structures ($\Delta$V$_{\rm LOS}$) in MCs with double and 
    triple $^{13}$CO structures. A single MC with double $^{13}$CO structures has one $\Delta$V$_{\rm LOS}$. 
    A single MC with triple $^{13}$CO structures have three $\Delta$V$_{\rm LOS}$ between each two $^{13}$CO structures or 
    two $\Delta$V$_{\rm MST}$ between each pair of $^{13}$CO structures connected by the minimal spanning tree (MST) algorithm. \label{fig:f_rel}}
\end{figure*}

\subsection{Molecular clouds with double $^{13}$CO structures} 
Each MC with double $^{13}$CO structures has a single $\Delta$V$_{\rm LOS}$, as illustrated in Figure \ref{fig:f_rel}, 
making it easier to analyse the relative movement between $^{13}$CO structures within a cloud. 
Here we focus on the 443 MC samples with double $^{13}$CO structures, whose physical properties 
are shown in Figures \ref{fig:cloud_paran} and \ref{fig:cloud_paraf}. 
Based on the kinematical distances of these clouds, 
the interquartile ranges of their physical scales are 1 -- 2 pc for those in the near group and 3.5 -- 5.6 pc for the far group.

\begin{figure*}[ht]
    \plotone{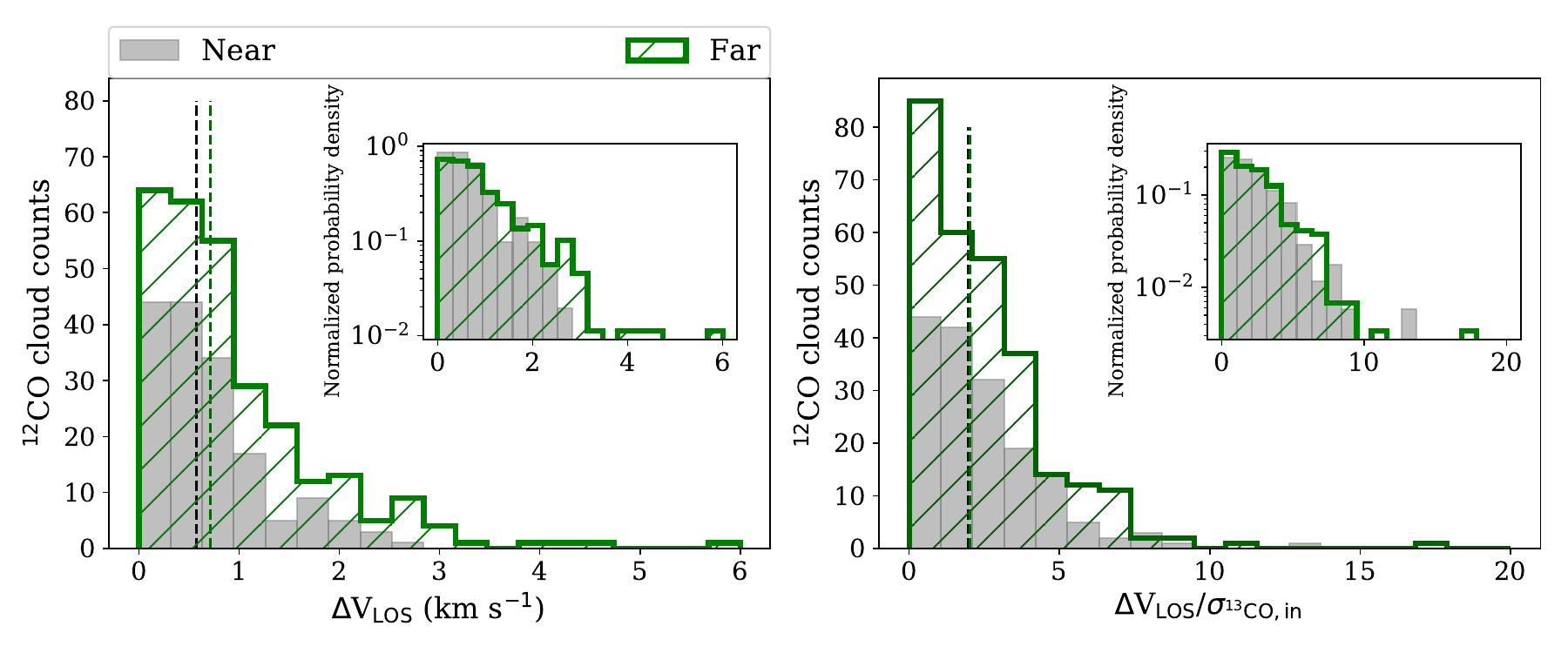} 
    \caption{Number distributions of the relative velocities ($\Delta$V$_{\rm LOS}$) and their ratios 
    with internal velocity dispersions of $^{13}$CO structures ($\Delta$V$_{\rm LOS}$/$\sigma_{\rm ^{13}CO,in}$) 
    for MCs with double $^{13}$CO structures. 
    The `Near' represents MCs have central velocities ranging from -30 to 25 km s$^{-1}$, 
    the `Far' means clouds with central velocities in a range of -95 to -30 km s$^{-1}$. 
    The vertical-dashed lines denote the corresponding median values. 
    In the upper-right corner of each panel, the corresponding normalized probability densities are presented.
    The normalized probability densities are presented as log scales, and the values in x-axis are binned as linear scales.\label{fig:f_cloud2}}
\end{figure*}

Figure \ref{fig:f_cloud2} presents the distributions of $\Delta$V$_{\rm LOS}$ and $\Delta$V$_{\rm LOS}$/$\sigma_{\rm ^{13}CO,in}$ ratios for these MCs. 
The quantiles at 0.05, 0.25, 0.5, 0.75, and 0.95 and the mean values 
of the $\Delta$V$_{\rm LOS}$ and $\Delta$V$_{\rm LOS}$/$\sigma_{\rm ^{13}CO,in}$ are listed in Tables \ref{tab:t_vel} and \ref{tab:t_dis}, respectively. 
Approximately 70$\%$ of $\Delta$V$_{\rm LOS}$ in the `Double' samples are less than 1 km s$^{-1}$, 
with less than 10$\%$ having values greater than 2 km s$^{-1}$ and reaching a maximal value of approximately 5 km s$^{-1}$. 
The distributions of $\Delta$V$_{\rm LOS}$ in the near and far groups have similar patterns,  
but the greater $\Delta$V$_{\rm LOS}$ (3 -- 5 km s$^{-1}$) have slightly higher probabilities to be observed in the far group. 
The quantiles of $\Delta$V$_{\rm LOS}$ in the far group are $\sim$ 1.3 times greater than those in the near group, 
probably due to the beam dilution effects on clouds with different distances, 
as previously discussed in Section 4.1 in Paper III, 
which also cause the linear separations among $^{13}$CO structures in the far group to be $\sim$ 3 times those in the near group.

For the distribution of $\Delta$V$_{\rm LOS}$/$\sigma_{\rm ^{13}CO,in}$, 
approximately 50$\%$ of values are less than 2, and around 70$\%$ are less than 3. 
About 10$\%$ of samples exhibit $\Delta$V$_{\rm LOS}$/$\sigma_{\rm ^{13}CO,in}$ greater than 5, with several samples reaching $\sim$ 12.  
The ratio values ($\Delta$V$_{\rm LOS}$/$\sigma_{\rm ^{13}CO,in}$) and their quantiles for the samples in the near and 
far groups display consistent distributions.   
Among these clouds, about 42$\%$ of $\Delta$V$_{\rm LOS}$ are identified as distinct velocity discontinuities ($\Delta V_{\rm dis}$), 
where $\Delta V_{\rm dis} > \sqrt{8\rm ln2}\sigma_{\rm ^{13}CO,in}$, as listed in Table \ref{tab:t_fac}. 
We also visually examine the velocity fields for these clouds with $\Delta V_{\rm dis}$, 
whose velocity fields are derived from the first moment of the $^{12}$CO and $^{13}$CO line emission, 
and present their averaged spectral lines of $^{12}$CO and $^{13}$CO line emission. 
The velocity fields of these clouds exhibit the following features: (1) relatively discontinuities on velocity fields, 
with one $^{13}$CO structure displaying blueshifted velocity and the other redshifted; 
(2) the averaged $^{13}$CO spectral lines resolving into two velocity components, as shown in several cases in Figure \ref{fig:f_2str}. 
Based on these observations, about 42$\%$ of MCs in the `double' regime show the signs of ongoing mergers or splits between clouds.
  
\subsection{Molecular clouds with triple $^{13}$CO structures}
As mentioned above, approximately 20$\%$ of the whole MC samples contain more than two $^{13}$CO structures. 
The analysis of relative velocities between multiple $^{13}$CO structures becomes increasingly complex 
as the number of $^{13}$CO structures increases. 
Here we focus on the 185 MC samples with triple $^{13}$CO structures, whose physical properties are presented in Figures \ref{fig:cloud_paran} and \ref{fig:cloud_paraf}. 
Based on the kinematical distances of these clouds, 
the interquartile ranges of their physical scales are 1.5 -- 2.5 pc for those in the near group and 4.7 -- 7.0 pc for the far group.

\subsubsection{Relative velocities between each two $^{13}$CO structures}
\begin{figure*}[ht]
    \plotone{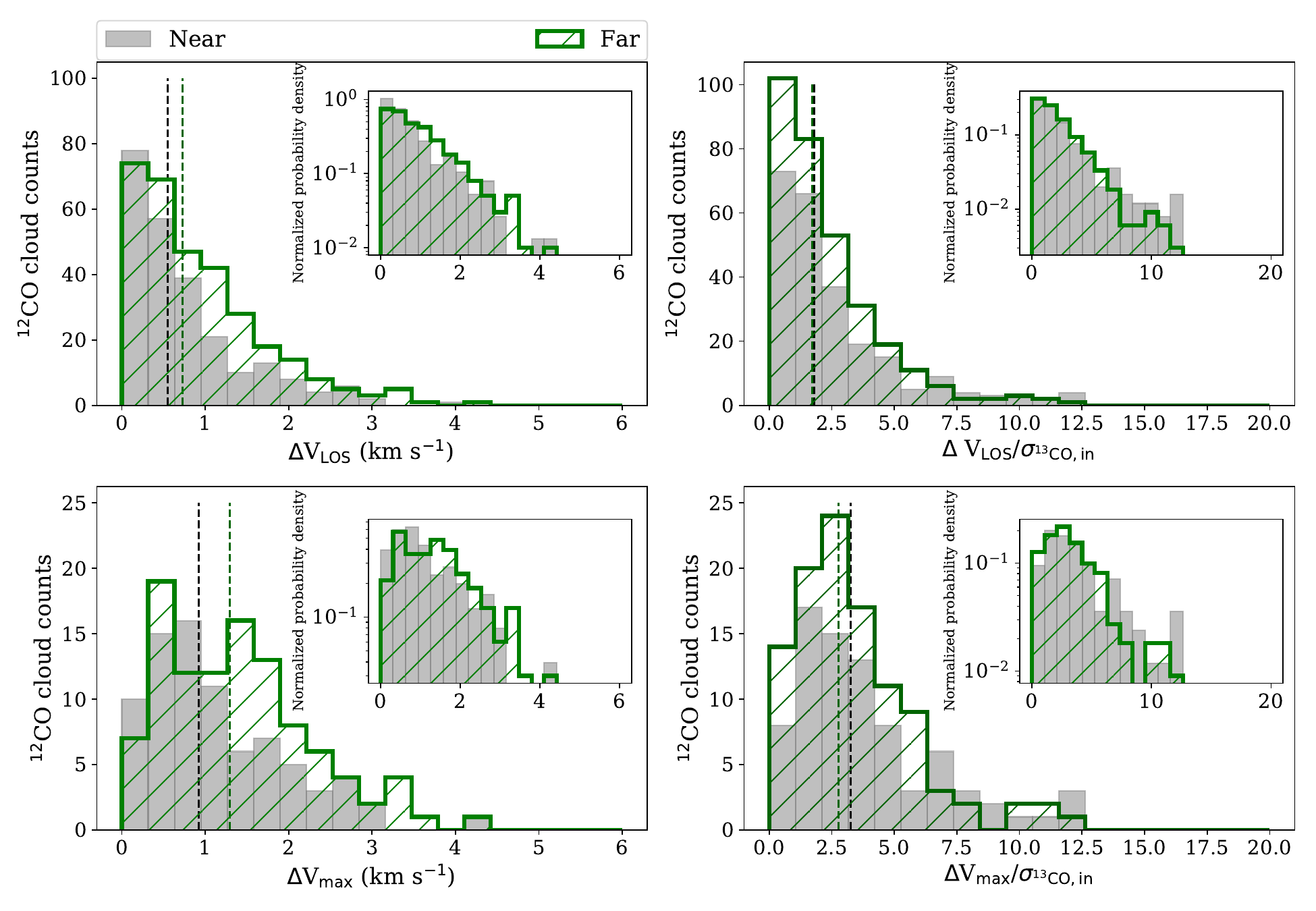} 
    \caption{\textbf{Upper panel}: number distributions of relative velocities ($\Delta$V$_{\rm LOS}$) between each two 
    $^{13}$CO structures in the MC samples having triple $^{13}$CO structures and 
    their ratios with internal velocity dispersions of $^{13}$CO structures ($\Delta$V$_{\rm LOS}$/$\sigma_{\rm ^{13}CO,in}$). 
    \textbf{Lower panel}: number distributions of the maximum relative velocity ($\Delta$V$_{\rm max}$) within each MC 
    and their ratios with internal velocity dispersions ($\Delta$V$_{\rm max}$/$\sigma_{\rm ^{13}CO,in}$). 
    The vertical-dashed lines show their median values. In the upper-right corner of each panel, 
    the corresponding normalized probability densities are presented.
    The normalized probability densities are presented as log scales, and the values in x-axis are binned as linear scales.\label{fig:f_cloud3}}
\end{figure*}

An individual MC with triple $^{13}$CO structures contains three $\Delta$V$_{\rm LOS}$ values between each two $^{13}$CO structures, 
as illustrated in Figure \ref{fig:f_rel}.
The distributions of the $\Delta$V$_{\rm LOS}$ values within 
these MC samples in both near and far groups are presented in Figure \ref{fig:f_cloud3}. 
The distributions of their maximum $\Delta$V$_{\rm LOS}$ ($\Delta$V$_{\rm max}$) within each sample are also presented, 
along with the quantiles at 0.05, 0.25, 0.5, 0.75, and 0.95, and the mean values 
of the $\Delta$V$_{\rm LOS}$ and $\Delta$V$_{\rm max}$, which are tabulated in Table \ref{tab:t_vel}. 
We find that the distributions of $\Delta$V$_{\rm LOS}$ in the MCs in the `triple' and `double' regimes are similar, 
and their quantiles and mean values are also close for both near and far groups. 
However, the $\Delta$V$_{\rm max}$, whose median value is about 0.92 km s$^{-1}$ in the near group 
and 1.3 km s$^{-1}$ in the far group, is significantly larger than those values for MCs in the `double' regime. 
It should be noted that $\Delta$V$_{\rm max}$ is the maximum value for the $\Delta$V$_{\rm LOS}$ between each pair of the $^{13}$CO structures. 
Therefore, the larger $\Delta$V$_{\rm max}$ is likely due to statistical probabilities for the larger number of the $^{13}$CO structure pairs in the `triple' regime. 

Figure \ref{fig:f_cloud3} displays distribution of the ratios between the $\Delta$V$_{\rm LOS}$ with $\sigma_{\rm ^{13}CO,in}$ and 
the ratios between $\Delta$V$_{\rm max}$ and $\sigma_{\rm ^{13}CO,in}$ for MCs with triple $^{13}$CO structures. 
In Table \ref{tab:t_dis}, the quantiles at 0.05, 0.25, 0.5, 0.75, and 0.95, as well as the mean values 
of these $\Delta$V$_{\rm LOS}$/$\sigma_{\rm ^{13}CO,in}$ and $\Delta$V$_{\rm max}$/$\sigma_{\rm ^{13}CO,in}$ values, are also listed.  
The distributions of $\Delta$V$_{\rm LOS}$/$\sigma_{\rm ^{13}CO,in}$ for MCs with triple $^{13}$CO structures are remarkably similar to those for MCs with double $^{13}$CO structures, 
with a nearly consistent interquartile range of $\sim$ 0.9 -- 3.4 and a median value of $\sim$ 2, in either the near or far group. 
However, the $\Delta$V$_{\rm max}$/$\sigma_{\rm ^{13}CO,in}$ values in the near group (median value of 3.26) and far group (median value of 2.77) 
are significantly greater than those values in MCs with double $^{13}$CO structures, which can be attributed to the greater $\Delta$V$_{\rm max}$. 
Overall, the distributions of the $\Delta$V$_{\rm LOS}$ for both `double' and `triple' regime samples are similar, 
as well as their $\Delta$V$_{\rm LOS}$/$\sigma_{\rm ^{13}CO,in}$ distributions. 
Such similarities suggest that the relative motions of $^{13}$CO structures within clouds are regulated 
by the fundamental processes that are independent of structure complexities and cloud scales. 

\begin{deluxetable*}{lcccccccc}
    \tablecaption{Radial relative velcotities ($\Delta$V$_{\rm LOS}$) between $^{13}$CO structures within $^{12}$CO MC samples. \label{tab:t_vel}}
    \tablewidth{0pt}
    \tablehead{\colhead{Groups} & \colhead{MC samples} & \colhead{Relative velocities} & \colhead{0.05} & \colhead{0.25} & \colhead{0.5} & \colhead{0.75} & \colhead{0.95} & \colhead{Mean}}
    \startdata
    \multirow{3}{*}{Near} & {MCs (Double)} & $\Delta$V$_{\rm LOS}$  & 0.05 & 0.28 & 0.57 & 0.95 & 1.93 & 0.75 \\\cline{2-9}
                          & \multirow{2}{*}{MCs (Triple)} & $\Delta$V$_{\rm LOS}$ & 0.05 & 0.23 & 0.55 & 1.0 & 2.47 & 0.77 \\
                          &                               & $\Delta$V$_{\rm max}$ & 0.15 & 0.53 & 0.92 & 1.69 & 2.78 & 1.15 \\\hline
    \multirow{3}{*}{Far}  & {MCs (Double)} & $\Delta$V$_{\rm LOS}$ & 0.08 & 0.35 & 0.71 & 1.24 & 2.64 & 0.94 \\\cline{2-9}
                          & \multirow{2}{*}{MCs (Triple)} & $\Delta$V$_{\rm LOS}$ & 0.08 & 0.33 & 0.73 & 1.29 & 2.47 & 0.93 \\
                          &                               & $\Delta$V$_{\rm max}$ & 0.29 & 0.67 & 1.3 & 1.89 & 3.14 & 1.39 \\\hline 
    \enddata
    \tablecomments{The quantiles at 0.05, 0.25, 0.5, 0.75, and 0.95 for the $\Delta$V$_{\rm LOS}$ (km s$^{-1}$) in their sequential data.
    The `Double' represents the 443 MCs with double $^{13}$CO structures and the `Triple' 
    corresponds to the 185 MCs having three $^{13}$CO structures. 
    The `Near’ and `Far’ represent the MC samples in the near and far groups, respectively. 
    The $\Delta$V$_{\rm LOS}$ is the relative velocity on the line of sight between $^{13}$CO structures. 
    There is one $\Delta$V$_{\rm LOS}$ in each `Double' sample and three $\Delta$V$_{\rm LOS}$ in each `Triple' sample. 
    The `$\Delta$V$_{\rm max}$' means the maximum value in the three $\Delta$V$_{\rm LOS}$ of each `Triple' sample. }
\end{deluxetable*}

\begin{deluxetable*}{lcccccccc}
    \tablecaption{The ratios between relative velcotities and internal velocity dispersions ($\Delta$V$_{\rm LOS}$/$\sigma_{\rm ^{13}CO, in}$) of $^{13}$CO structures within $^{12}$CO MC samples. \label{tab:t_dis}}
    \tablewidth{0pt}
    \tablehead{\colhead{Groups} & \colhead{MC samples} & \colhead{Ratios} & \colhead{0.05} & \colhead{0.25} & \colhead{0.5} & \colhead{0.75} & \colhead{0.95} & \colhead{Mean}}
    \startdata
    \multirow{3}{*}{Near} & {MCs (Double)} & $\Delta$V$_{\rm LOS}$/$\sigma_{\rm ^{13}CO, in}$ & 0.28 & 0.97 & 1.96 & 3.41 & 5.93 & 2.43 \\\cline{2-9}
                          & \multirow{2}{*}{MCs (Triple)} & $\Delta$V$_{\rm LOS}$/$\sigma_{\rm ^{13}CO, in}$ & 0.18 & 0.93 & 1.8 & 3.44 & 8.07 & 2.6 \\
                          &                               & $\Delta$V$_{\rm max}$/$\sigma_{\rm ^{13}CO, in}$ & 0.88 & 1.89 & 3.26 & 4.89 & 9.63 & 3.89 \\\hline
    \multirow{3}{*}{Far}  & {MCs (Double)} & $\Delta$V$_{\rm LOS}$/$\sigma_{\rm ^{13}CO, in}$ & 0.24 & 0.88 & 2.04 & 3.34 & 6.46 & 2.45 \\\cline{2-9}
                          & \multirow{2}{*}{MCs (Triple)} & $\Delta$V$_{\rm LOS}$/$\sigma_{\rm ^{13}CO, in}$ & 0.22 & 0.77 & 1.72 & 3.09 & 6.2 & 2.29 \\
                          &                               & $\Delta$V$_{\rm max}$/$\sigma_{\rm ^{13}CO, in}$ & 0.62 & 1.85 & 2.77 & 4.69 & 7.97 & 3.44 \\\hline 
    \enddata
    \tablecomments{The quantiles at 0.05, 0.25, 0.5, 0.75, and 0.95 for the $\Delta$V$_{\rm LOS}$/$\sigma_{\rm ^{13}CO, in}$ and $\Delta$V$_{\rm max}$/$\sigma_{\rm ^{13}CO,in}$ 
    in their sequential data. The MC samples are reported in Table \ref{tab:t_vel}.}
\end{deluxetable*}

\subsubsection{Relative velocities between $^{13}$CO structures connected by MST}
For a cloud containing three $^{13}$CO structures, 
the relative velocities between each pair of them do not take the spatial distribution of $^{13}$CO structures into account. 
In order to examine whether the relative motion between $^{13}$CO structures is primarily random motion or systematic motion, 
we connect the centroid coordinates of $^{13}$CO structures together using the minimal spanning tree (MST), 
which minimizes the sum of the angular separation between $^{13}$CO structures. 
Furthermore, we analyzed the relative velocities ($\Delta$V$_{\rm MST}$) between the connected $^{13}$CO structures. 
There are two $\Delta$V$_{\rm MST}$ values in each MC with triple $^{13}$CO structures, as illustrated in Figure \ref{fig:f_rel}. 

\begin{figure*}[ht]
    \plotone{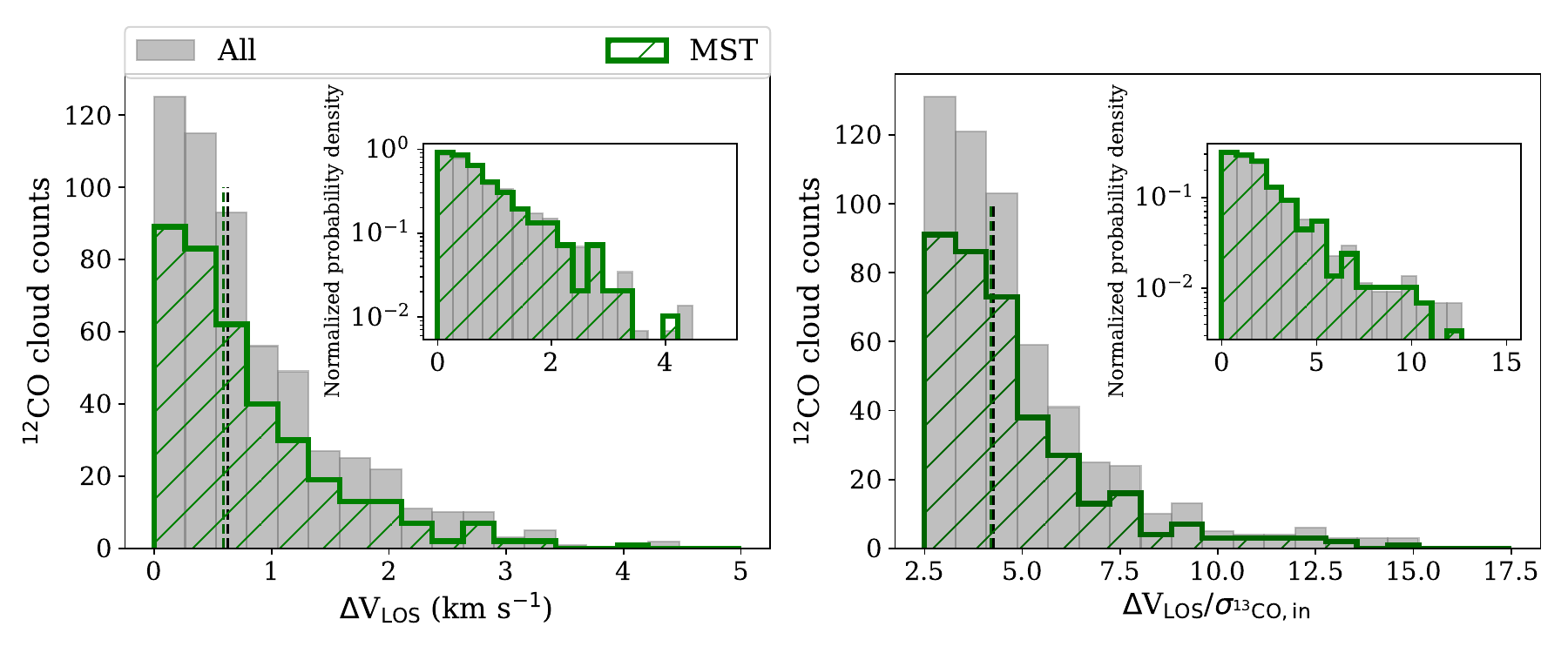} 
    \caption{The distribution of relative velocities between $^{13}$CO structures for 185 MCs in the `triple' regime. 
    The gray histograms represent the whole 555 $\Delta$V$_{\rm LOS}$ from the arbitrary two of $^{13}$CO structures within 185 samples. 
    The green histograms mean the 370 $\Delta$V$_{\rm MST}$ between the $^{13}$CO structures connected through the minimal spanning three (MST) algorithm, 
    there are two $\Delta$V$_{\rm MST}$ within a cloud. The corresponding vertical-dashed lines show their median values. 
    In the upper-right corner of each panel, the corresponding normalized probability densities are presented.
    The normalized probability densities are presented as log scales, and the values in x-axis are binned as linear scales. \label{fig:f_mst}}
\end{figure*}

The distributions of $\Delta$V$_{\rm LOS}$ and $\Delta$V$_{\rm MST}$ for the 185 MCs in the triple regime, 
are showed in Figure \ref{fig:f_mst}. 
We find that their distributions are nearly accordant, especially for their normalized probability densities. 
Such similarity means the relative motions between $^{13}$CO structures are random and regulated by fundamental processes. 
Turbulent flows are a promising candidate for this process, as they are driven by dynamical processes in different scales, 
including galactic differential rotation and shear \citep{Kim2006, Bonnell2006, Dobbs2008}, 
large-scale instabilities \citep{Wada2002, Tasker2009}, and stellar feedback from the supernova explosions 
\citep{Brunt2009, Skarbinski2023, Watkins2023} and HII regions \citep{Silk1985, Krumholz2006}. 

\subsubsection{Distinct velocity discontinuities within MCs} 

Among 185 MCs with triple $^{13}$CO structures, 21.1$\%$ of these MCs have one $\Delta V_{\rm dis}$, 
33.5$\%$ of samples exhibit two $\Delta V_{\rm dis}$, and 7$\%$ of them display three $\Delta V_{\rm dis}$, as listed in Table \ref{tab:t_fac}.
We further conducted visual inspections of the velocity fields (first moment maps of CO emission) of these clouds 
and the averaged spectral lines of their $^{12}$CO and $^{13}$CO emissions. 
In Figure \ref{fig:f_3str1}, we present the velocity fields for clouds with two $\Delta V_{\rm dis}$, 
which have two $^{13}$CO structures with redshifted (blueshifted) velocities, and 
the other $^{13}$CO structure exhibits blueshifted (redshifted) velocities. 
Furthermore, the averaged spectra of $^{13}$CO emission are resolved into two velocity components. 
Figure \ref{fig:f_3str2} shows the velocity fields for clouds with three $\Delta V_{\rm dis}$, 
featuring three different velocity fields, each harboring one $^{13}$CO structure. 
The averaged spectra of $^{13}$CO emission are also decomposed into three velocity components.

\begin{figure*}[ht]
    \plotone{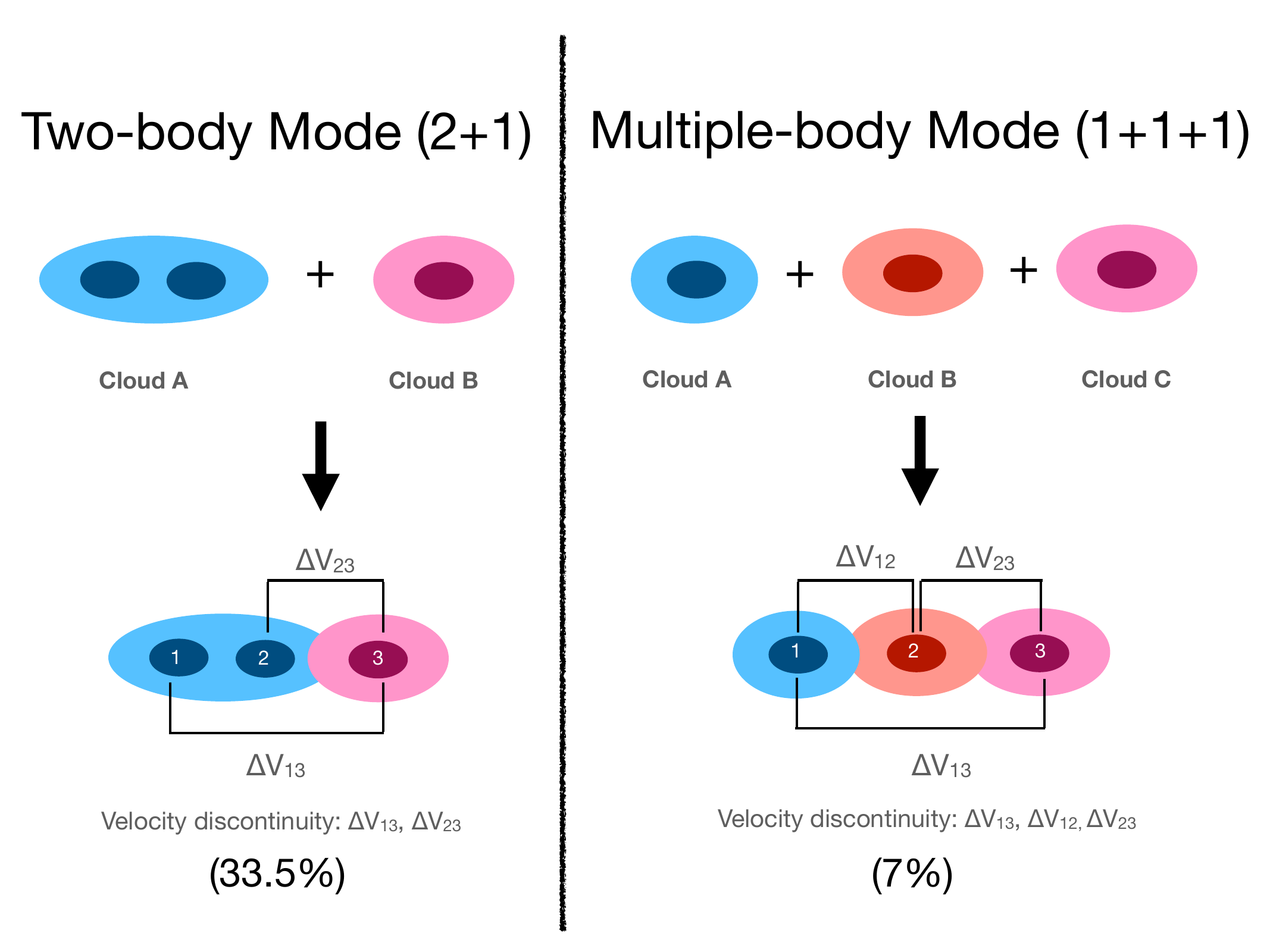} 
    \caption{Schematic illustrations of the two kinds of combination modes for MCs with triple $^{13}$CO structures. \label{fig:f_merger}}
\end{figure*}

We suggest that the build-up processes for MCs containing triple $^{13}$CO structures could be a result of two clouds combination, 
one with double $^{13}$CO structures and the other with single $^{13}$CO structure (two-body mode), 
or the assembly of three clouds, each with a single $^{13}$CO structure (multiple-body mode). 
Figure \ref{fig:f_merger} illustrates the velocity fields of cloud interaction through two-body and multiple-body modes, respectively.   
In the two-body mode, one $^{13}$CO structure exhibits redshifted (blueshifted) velocity, 
while the other two $^{13}$CO structures display the blueshifted (redshifted) velocity,
with two $\Delta V_{\rm dis}$ between one $^{13}$CO structure and the other two $^{13}$CO structures within clouds. 
In the multiple-body mode, each $^{13}$CO structure has a distinct velocity, resulting in three discontiguous velocity fields in the cloud's velocity fields.
Thus three $\Delta V_{\rm dis}$ between each pair of $^{13}$CO structures are expected within the cloud. 
That is, the velocity fields of clouds are constructed by velocities from these $^{13}$CO structures and exhibit three discontiguous velocities. 

Thus, in the `triple' regime, about 40.5$\%$ of MCs show distinct velocity discontinuities, 
with 33.5$\%$ of clouds displaying the signatures of the two-body mode and only 7$\%$ of clouds presenting the multiple-body mode. 
This suggests that cloud mergers or splits tend to occur between two MCs. 
\cite{Horie2023} simulated the fraction of mass from each progenitor within a colliding GMC and 
found the sum of the two most significant fractions is mostly close to 1, 
also suggesting a two-body mode for the majority of MC collisions. 
We should note that 21.1$\%$ of MCs have one $\Delta V_{\rm dis}$ between the arbitrary two $^{13}$CO structures, 
but do not exhibit the distinct discontinuity in velocity field caused by the bulk motions of $^{13}$CO structures. 
These clouds are not included into the MCs in triple regime showing distinct velocity discontinuities. 

In summary, the similarity of the $\Delta$V$_{\rm LOS}$ distribution in the double and triple regimes  
indicates the relative motion between $^{13}$CO structures is random and independent of cloud scales. 
Additionally, a considerable portion of distinct velocity discontinuities between $^{13}$CO structures ($\Delta V_{\rm dis}$) 
is observed within $^{12}$CO MCs.  
These results are coincident with our previous findings on regularly spaced $^{13}$CO structures within MCs (Paper III) and 
distinctly demonstrate on how the relative motion of $^{13}$CO structures provides a dominant contribution to 
the kinetic energy of MCs (Paper IV). 
Thus these results further support our previously proposed picture that $^{13}$CO structures 
act as the building blocks of MCs, and the transient processes of MCs proceed by slow mergers or splits among 
these fundamental blocks.

\begin{deluxetable}{lcccc}
    \tablecaption{Fractions of MC samples with the distinct velocity discontinuities. \label{tab:t_fac}}
    \tablewidth{0pt}
    \tablehead{\colhead{MC samples} & \colhead{Zero} & \colhead{One} &
    \colhead{Two} & \colhead{Three}\\}
    \startdata
    MCs (Double) & 58.2$\%$ & 41.8$\%$ & $\text{- -}$ &  $\text{- -}$ \\
    MCs (Triple) & 38.4$\%$ & 21.1$\%$ & 33.5$\%$ & 7.0$\%$ \\
    \enddata
    \tablecomments{The fractions of MCs not having the distinct velocity discontinuity ($\Delta V_{\rm dis}$) and 
    having one $\Delta V_{\rm dis}$ in MC samples with double $^{13}$CO structures. 
    Also, the fractions of MCs not having $\Delta V_{\rm dis}$ (zero) and having 
    one, two, and three $\Delta V_{\rm dis}$ in MC samples with triple $^{13}$CO structures.}
\end{deluxetable}

\section{Discussion}

\subsection{Observational bias}

Our observed MC samples are identified as the contiguous structures in the position-position-velocity (PPV) space, 
which have $^{12}$CO(1-0) emission intensities above a certain threshold. 
However, it is important to note that the structure identification in the PPV intensity structures, 
compared with the real structures in the position-position-position (PPP), may have some bias due to the effects of projection. 
This bias is attributed to two main factors. Firstly, distinct structures in the PPP space that 
move at similar velocities along the line of sight can be projected as a contiguous structure in the PPV space.  
Secondly, a single structure in the PPP space that has distinct velocity gradients can result in multiple PPV structures. 
Unfortunately, these ambiguities that are caused by projection cannot be directly resolved through a simple method in observations. 
To mitigate this issue, we combine the observed and simulated results.  

\begin{figure}[ht]
    \plotone{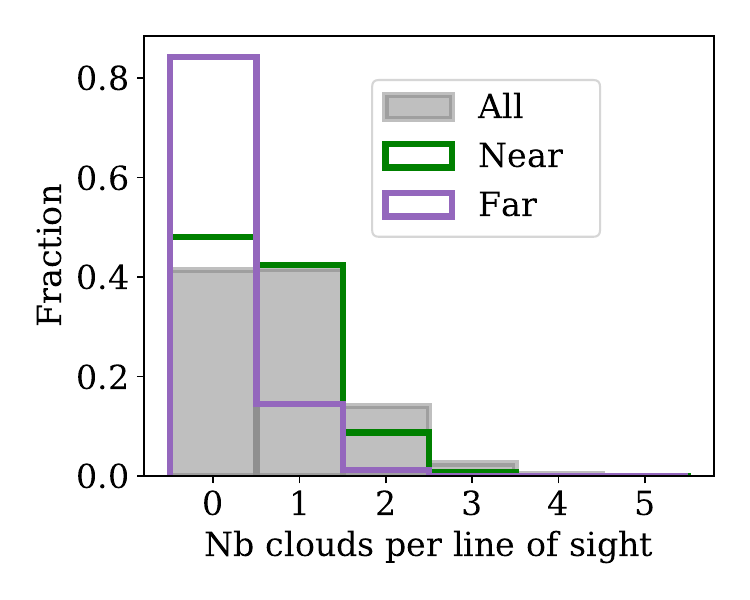} 
    \caption{Histogram of the number of clouds along line of sight in each pixel. 
    The y-axis is given in fractions of the total number of pixels in the region of 104$^{\circ}$.75 $< l <$ 150$^{\circ}$.25, $|b| < 5^{\circ}.25$. 
    The `All' means the line of sight with $-$95 $<$ V$_{\rm LSR}$ $<$ 25 km s$^{-1}$, 
    the `Near' is along the line of sight with $-$30 $<$ V$_{\rm LSR}$ $<$ 25 km s$^{-1}$, 
    and the `Far' is in a range of $-$95 $<$ V$_{\rm LSR}$ $<$ $-$30 km s$^{-1}$. \label{fig:f_cloudN}}
\end{figure}

In observations, the number of velocity components per spectrum along the line of sight can give us 
a global insight into the level of velocity crowding in the observed region of sky. 
Our MC samples are located in the second Galactic quadrant with l = 104$^{\circ}$.75 -- 150$^{\circ}$.25, 
$|b| < 5^{\circ}.25$, and $-$95 km s$^{-1}$  $<$ V$_{\rm LSR}$ $<$ 25 km s$^{-1}$. 
In total, we have extracted 18,190 $^{12}$CO clouds in the PPV space of $^{12}$CO emission within this region.
Figure \ref{fig:f_cloudN} shows the number distribution of 18,190 $^{12}$CO clouds per line of sight. 
We find that on $\sim$ 40$\%$ of this sky region, there is no cloud on the line of sight. 
In addition, on another $\sim$ 40$\%$ of the region, there is only one cloud along the line of sight. 
In approximately 15$\%$ of this area, two clouds overlap on the line of sight, 
and $\sim$ 5$\%$ of positions intersect three or four clouds. 
Furthermore, we count the number of molecular clouds along the line of sight, 
which are in the velocity range of (-95 -30) km s$^{-1}$ (Far) and (-30 25) km s$^{-1}$ (Near), respectively. 
As shown in Figure \ref{fig:f_cloudN}, approximately 50$\%$ of positions have no cloud and $\sim$ 40$\%$ have only one cloud 
on the line of sight in the near range. However, on the line of sight in the far range, 
$\sim$ 85$\%$ of areas have no cloud and $\sim$ 15$\%$ have only one cloud. 
Additionally, \cite{Riener2020, Miville2017} also have revealed that in most of the sky, apart from the vicinity of the Galactic center, 
there are only single or double velocity components along the line of sight.  
Considering that $\sim$ 10$\%$ of positions are covered by more than one cloud in the near velocities range of (-30 25) km s$^{-1}$, 
the likelihood for our clouds, 95$\%$ of which have $^{12}$CO velocity span less than 10 km s$^{-1}$, 
being projected together by multiple clouds on the line of sight is roughly 2$\%$. 

In numerical simulations, the relationship between the PPP density structures and PPV intensity structures has been explored in different aspects. 
In a simulated barred-spiral galaxy, \cite{Pan2015} found that $\sim$ 70$\%$ of clouds had single counterparts in both PPP and PPV data sets.  
At the MC's scale, \cite{Beaumont2013} suggested that structures traced in $^{12}$CO lines were more affected by 
overlap than those traced in $^{13}$CO lines due to the opacity of $^{12}$CO lines.   
The scaling relations, such as mass-size and size-linewidth, were found 
to be quite robust to projection by \cite{Ballesteros2002, Beaumont2013}.
In terms of velocity and density structures, 
\cite{Pichardo2000} reveal that the PPV features are more representative of the line of sight velocity than the density field.
\cite{Burkhart2013} further found that the dominant structures in the PPP and PPV are 
strictly linked to supersonic gas.  
For our MC samples, which are extracted in the contiguous $^{12}$CO emission in the PPV space of the Milky Way, 
we primarily focus on their internal velocity structures traced by the $^{13}$CO emission. 
Their velocity dispersions are greater than $\sim$ 0.5 km s$^{-1}$, as listed in Table 2 of \cite{Yuan2023b}. 
Comparing the sound speed c$_{s}$ = (kT$_{\rm kin}$/$\mu$m$_{\rm H}$)$^{1/2}$ $\sim$ 0.2 km s$^{-1}$ for gas kinetic temperature T$_{\rm kin}$=10 K and 
mean molecular weight $\mu$=2.33, thus these MC samples are thought to be supersonic. 
Therefore, based on these simulated results, 
the line of sight velocities of clouds are most likely represented by their velocity structures traced by the $^{13}$CO lines.

\subsection{Comparison with simulated works} 

The observed relative velocity between a pair of $^{13}$CO structures is  
the absolute difference between the observed centroid velocity of each $^{13}$CO structure, 
which is the three-dimensional (3D) velocity of each $^{13}$CO structure ($\vec{V}_{\rm 3D}$) projected along the line of sight. 
We determine an angle of $\theta_{\rm j}$ between the direction of $\vec{V}_{\rm 3D,j}$ at the $j$th $^{13}$CO structure and the line of sight, 
thus the observed $\Delta V_{\rm LOS}$ between $j$th and ($j$-1)th $^{13}$CO structure is $\Delta V_{\rm LOS} = \left|V_{\rm 3D,j} \cos \theta_{j} - V_{\rm 3D,j-1} \cos \theta_{j-1}\right|$. 
While the relative velocity in 3D ($\Delta \vec{V}_{\rm 3D}$) between two $^{13}$CO structures is $\Delta \vec{V}_{\rm 3D} = \vec{V}_{\rm 3D,j} - \vec{V}_{\rm 3D,j-1}$. 
We further assume a random distribution for the angle $\theta_{j}$ and get a mean value of 
$\langle \cos \theta_{\rm j} \rangle$ = 1/$\sqrt{2}$.  
Under these assumptions, we can roughly estimate the values of the relative velocities in 3D as 
$\Delta V_{\rm 3D}= \sqrt{2}\Delta V_{\rm LOS}$. 
Our results indicate that over 95$\%$ of the $\Delta V_{\rm 3D}$ are less than 5 km s$^{-1}$, 
regardless of whether they are in the double or triple regime. 

Based on our previous works, it appears that the distinct velocity discontinuities between $^{13}$CO structures 
are most likely a result of cloud merging or splitting.  
In order to gain further insight into cloud motion and its relation to the internal $^{13}$CO structure motion, 
it is necessary to systematically investigate cloud-cloud motion in further.
Numerical simulations of the cloud's motion can provide useful insight at this point.
For instance, the distribution of relative velocities for the cloud-cloud collisions or mergers (CCCs) has been analyzed in a simulated galaxy \citep{Horie2023}. 
The distribution of the cloud collision speeds exhibits a similar trend to that of the observed relative velocities between internal $^{13}$CO 
structures \citep{Horie2023}. However, the simulated values of cloud collision speeds are distributed at a peak of $\sim$ 7 km s$^{-1}$, with about 65$\%$ of 
them being less than 10 km s$^{-1}$ \citep{Horie2023}. 
Nonetheless, \cite{Skarbinski2023} have found that about 80$\%$ of mergers occur at a relative velocity of less than 5 km s$^{-1}$, 
implying that cloud mergers have a greater impact on aggregating mass into larger molecular complexes than 
generating shocks and further altering density structures of the merged cloud. 
This finding was also discussed in \cite{Dobbs2015, Jeffreson2021}.  
In observation, \cite{Fukui2021} summarized the recent observational results on about 65 high-mass star-forming regions triggered by CCCs 
and found that their median collision velocity is $\sim$ 5 km s$^{-1}$. 
The distribution of internal relative velocities between $^{13}$CO structures is fairly consistent with that for 
the merged speeds of clouds in these simulated works and observed results.

Furthermore, it is worth noting that the observed velocity discontinuities in this research are interpreted as the signatures of mergers or splits 
undergoing within MCs. However, we cannot exclude the possibilities of rotation motion, which likely arises 
from the turbulent flows from the shearing action of differential Galactic rotation \citep{Fleck1981, Belloche2013, Arroya2022}. 
At least, our observed results also provide a constraint for the velocity gradients on the rotation motion of clouds. 
Additionally, further analysis on the distribution of the angular momentum from cloud scales down to internal substructures are helpful 
in resolving the process of the angular momentum redistribution within MCs.        

\section{Conclusions}
Our study analyzes a sample of 443 $^{12}$CO MCs with double $^{13}$CO structures, and 185 $^{12}$CO MCs with triple $^{13}$CO structures 
from the MWISP CO survey. 
Our objective is to investigate the relative velocities on the line of sight between $^{13}$CO structures ($\Delta$V$_{\rm LOS}$) within $^{12}$CO clouds 
and identify the proportion of MCs exhibiting instinct velocity discontinuities. 
Our findings can be summarized as follows: 

1. Approximately 70$\%$ of $\Delta$V$_{\rm LOS}$ values are less than $\sim$ 1 km s$^{-1}$, 
and roughly 10$\%$ of values exceed 2 km s$^{-1}$ with the maximum value reaching $\sim$ 5 km s$^{-1}$. 
For the ratios between $\Delta$V$_{\rm LOS}$ values and internal velocity dispersions of $^{13}$CO structures ($\Delta$V$_{\rm LOS}$/$\sigma_{\rm ^{13}CO,in}$), 
approximately 70$\%$ of them are less than 3 and about 10$\%$ are greater than 5, with several samples reaching $\sim$ 12.

2. The distributions of $\Delta$V$_{\rm LOS}$ for MCs with double and triple $^{13}$CO structures are similar, 
as well as their $\Delta$V$_{\rm LOS}$/$\sigma_{\rm ^{13}CO,in}$. 
Additionally, in the triple regime, distributions of $\Delta$V$_{\rm LOS}$ between  
arbitrary two $^{13}$CO structures and pairs of $^{13}$CO structrues connected by the minimal spanning tree are also similar. 
Such similarities suggest that the relative motions of $^{13}$CO structures within clouds are random and 
regulated by the fundamental processes that are independent of structure complexities and cloud scales. 
Turbulent flows are a promising candidate, as they can be driven by dynamical processes in different scales.

3. About 40$\%$ of MCs in either double or triple regimes exhibit distinct velocity discontinuities, 
with relative velocities between $^{13}$CO structures greater than the internal linewidths of $^{13}$CO structures. 
The velocity fields for this portion of MC samples present redshifted and blueshifted velocities that come from the bulk motions of $^{13}$CO structures. 

4. Among the 40$\%$ of samples in the triple regime showing distinct velocity discontinuities, 
$\sim$ 33$\%$ exhibit the signatures of bulk motions through the two-body motion, 
whereas the remaining $\sim$ 7$\%$ show the feasures of combinations through the multiple-body motion. 
This suggests that cloud mergers or splits tend to occur between two MCs. 

5. The indication of random motion among $^{13}$CO structures and a considerable portion 
of distinct velocity discontinuities between $^{13}$CO structures with $^{12}$CO clouds provide 
further evidence for our previously proposed picture that $^{13}$CO structures act as the building blocks of MCs, 
and the transient processes of MCs proceed by slow mergers or splits among these fundamental blocks.

\begin{acknowledgments}
    This research made use of the data from the Milky Way Imaging Scroll Painting (MWISP) project, 
    which is a multi-line survey in $^{12}$CO/$^{13}$CO/C$^{18}$O along the northern galactic plane with PMO-13.7m telescope. 
    We are grateful to all of the members of the MWISP working group, particulaly the staff members at the PMO-13.7m telescope, 
    for their long-term support. LY acknowledges Haoran Feng for his support on some data analysis scripts. 
    This research was supported by the National Natural Science Foundation of China through grant 
    12303034 $\&$ 12041305 and the Natural Science Foundation of Jiangsu Province through grant BK20231104.  
    MWISP was sponsored by the National Key R\&D Program of China with grant 2023YFA1608000 $\&$ 2017YFA0402701 and 
    the CAS Key Research Program of Frontier Sciences with grant QYZDJ-SSW-SLH047. 
\end{acknowledgments}

\textbf{Data Availability.} The extracted $^{12}$CO line data for the 18,190 $^{12}$CO clouds and the extracted 
$^{13}$CO line data within the 2851 $^{12}$CO clouds are publicly available at DOI:\href{https://doi.org/10.57760/sciencedb.j00001.00427}{10.57760/sciencedb.j00001.00427}.

\software{Astropy \citep{astropy2013, astropy2018}, Scipy \citep{scipy2020}, Matplotlib \citep{Hunter2007}}
    
\clearpage
\appendix
\renewcommand{\thefigure}{\Alph{section}\arabic{figure}}
\renewcommand{\theHfigure}{\Alph{section}\arabic{figure}}
\setcounter{figure}{0}

\section{Basic parameters for molecular clouds with double and triple $^{13}$CO structures}
\begin{figure*}[h]
    \plotone{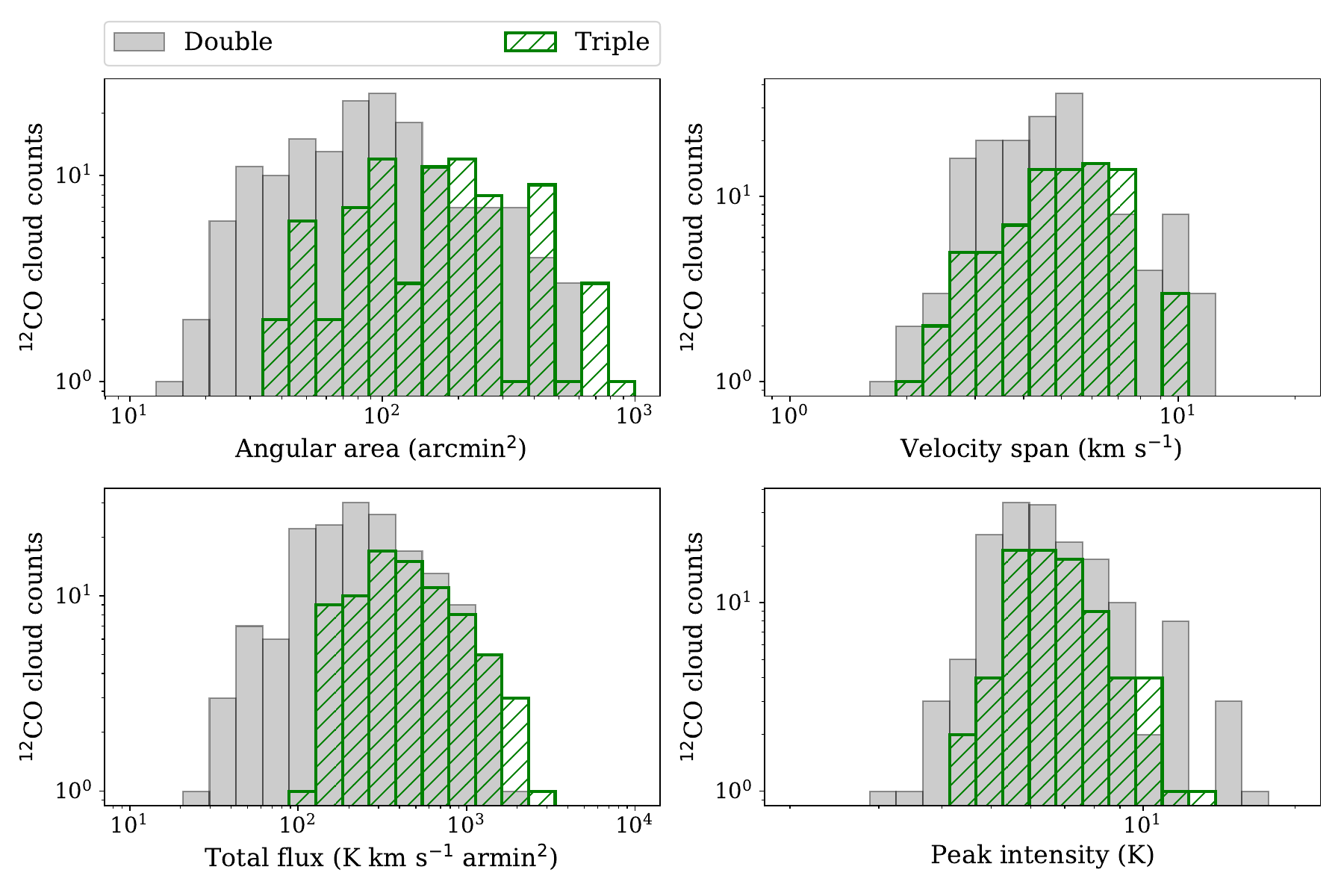}
    \caption{Number distributions of angular areas, velocity spans, integrated fluxes, and peak intensities of 
    $^{12}$CO line emission for MC samples in the near group. Most of these samples are located in the local arm and have kinematical 
    distances of $\sim$ 0.5 kpc. The gray histograms represent 163 $^{12}$CO MCs with double $^{13}$CO structures, 
    and the green histograms represent 80 $^{12}$CO MCs with triple $^{13}$CO structures. \label{fig:cloud_paran}}
\end{figure*}

\begin{figure*}[h]
    \plotone{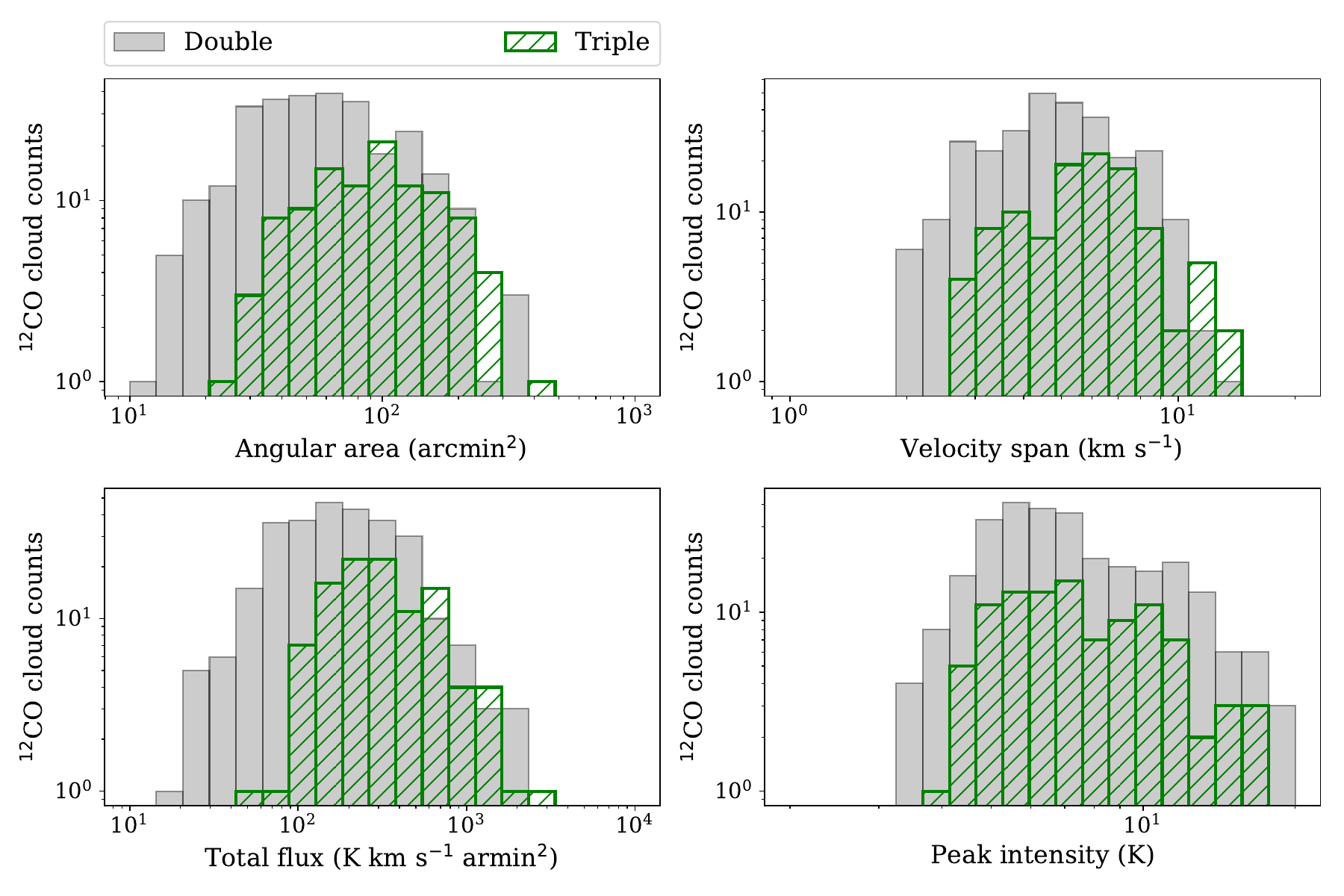}
    \caption{Number distribution of angular areas, velocity spans, integrated fluxes, and peak intensities of 
    $^{12}$CO line emission for MC samples in the far group, most of which are distributed in the Perseus arm with kinematical 
    distances of $\sim$ 2 kpc. The gray histograms represent 280 $^{12}$CO MCs with double $^{13}$CO structures, 
    and the green histograms represent 105 $^{12}$CO MCs with triple $^{13}$CO structures. \label{fig:cloud_paraf}}
\end{figure*}

Figures \ref{fig:cloud_paran} and \ref{fig:cloud_paraf} present distributions of 
angular areas, velocity spans, integrated fluxes, and peak intensities of $^{12}$CO line emission 
for MC samples in the near and far groups, respectively.

\renewcommand{\thefigure}{\Alph{section}\arabic{figure}}
\renewcommand{\theHfigure}{\Alph{section}\arabic{figure}}
\setcounter{figure}{0}

\section{Molecular clouds with distinct velocity discontinuities}
\begin{figure*}[ht]
    \plotone{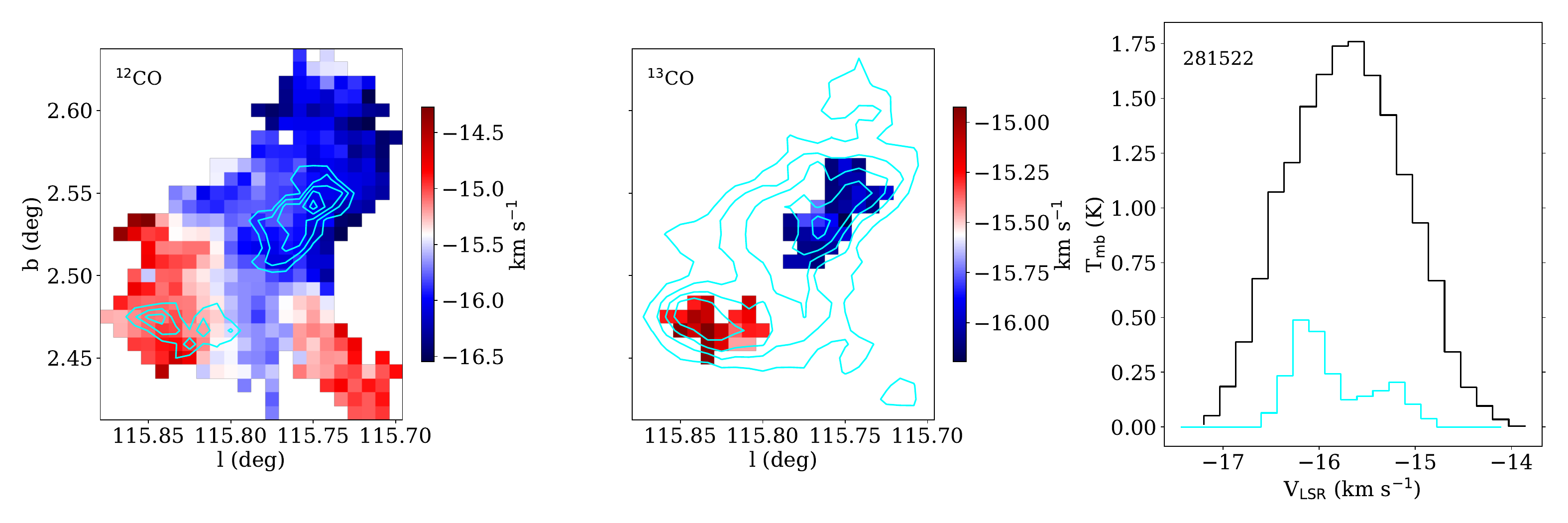}
    \plotone{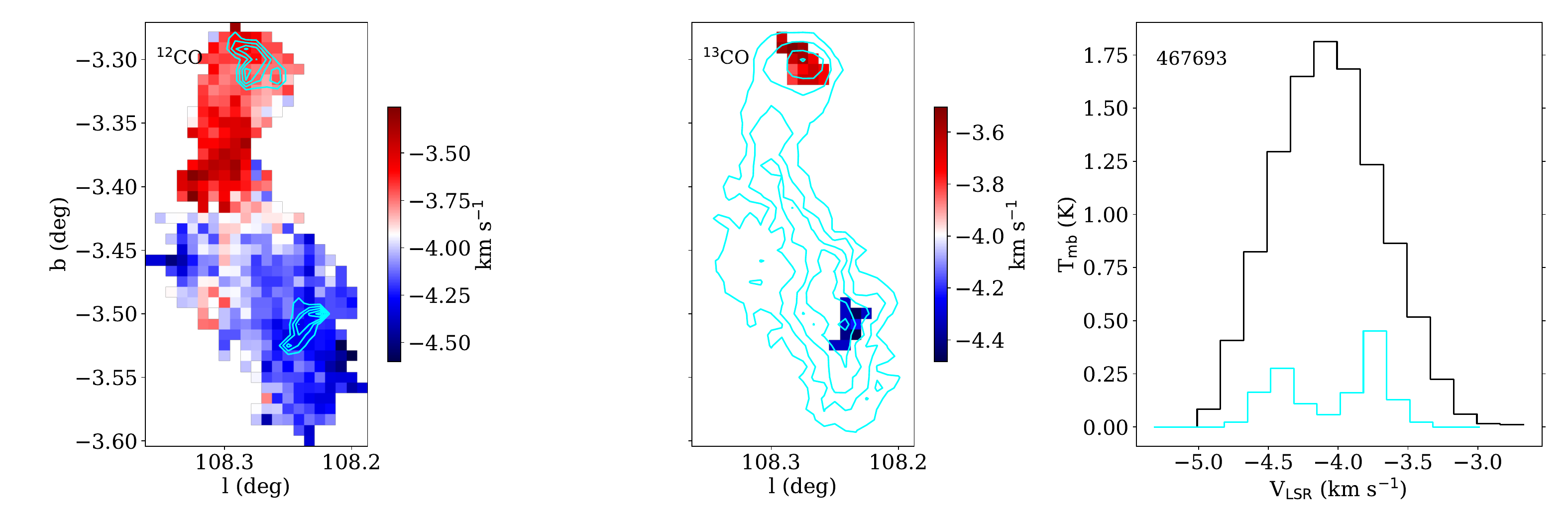}
    \plotone{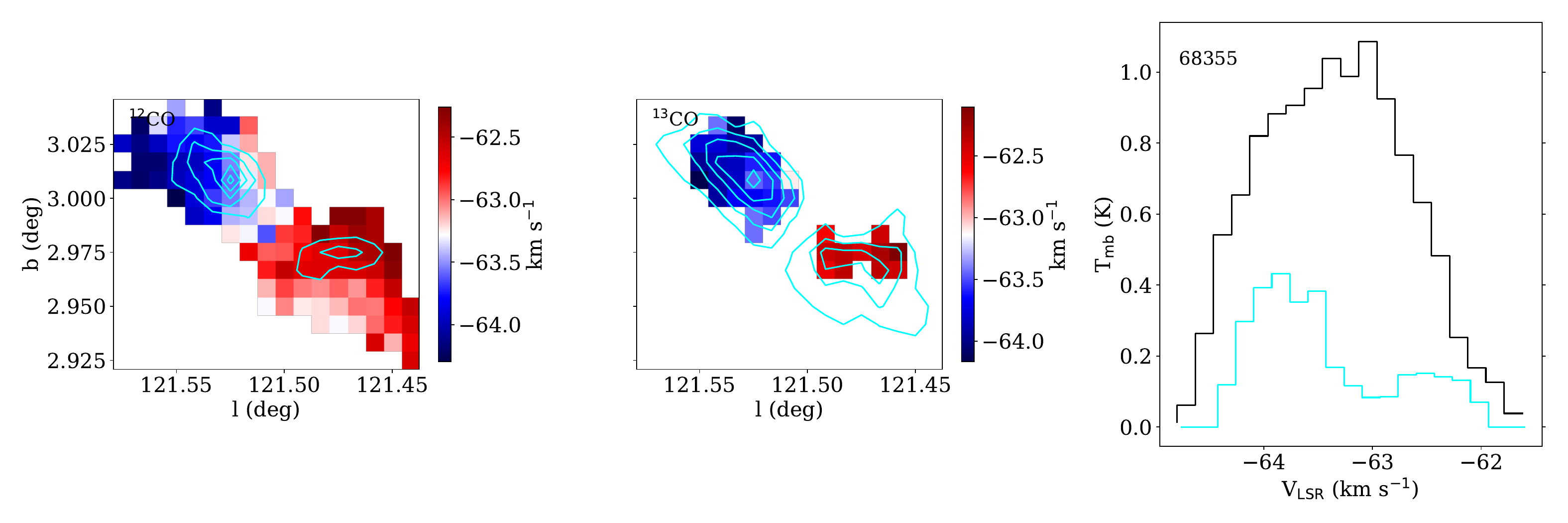}
    \plotone{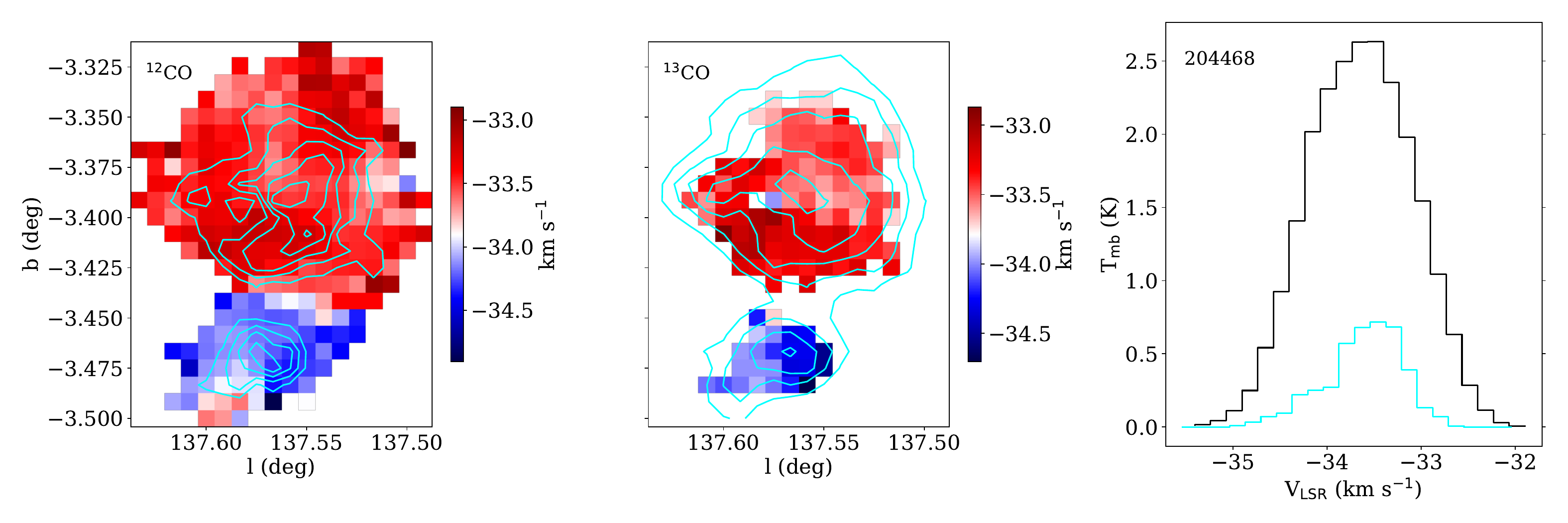}
    \caption{Velocity fields of MCs with double $^{13}$CO structures, whose relative velocities are characterized as 
    distinct velocity discontinuities. \textbf{Left panel}: the color map represents the first moment map (velocity field) of $^{12}$CO 
    emission of MCs, and the cyan contours show the moment zero map (velocity-integrated intensity) of $^{13}$CO line emission, ranging 
    from 10$\%$ to 90$\%$ in increments of 20$\%$ of its maximum value. \textbf{Middle panel}: the color map represents the first moment map 
    (velocity field) of $^{13}$CO emission of MCs, and the cyan contours show the moment zero map (velocity-integrated intensity) 
    of $^{12}$CO line emission, ranging from 10$\%$ to 90$\%$ in increments of 20$\%$ of its maximum value. \textbf{Right panel}: 
    the averaged spectral lines for the extracted $^{12}$CO line emission (black) and $^{13}$CO line emission (cyan) within MCs. The number noted in the 
    upper-left corner is the number ID for MC samples. \label{fig:f_2str}}
\end{figure*}

\begin{figure*}[ht]
    \plotone{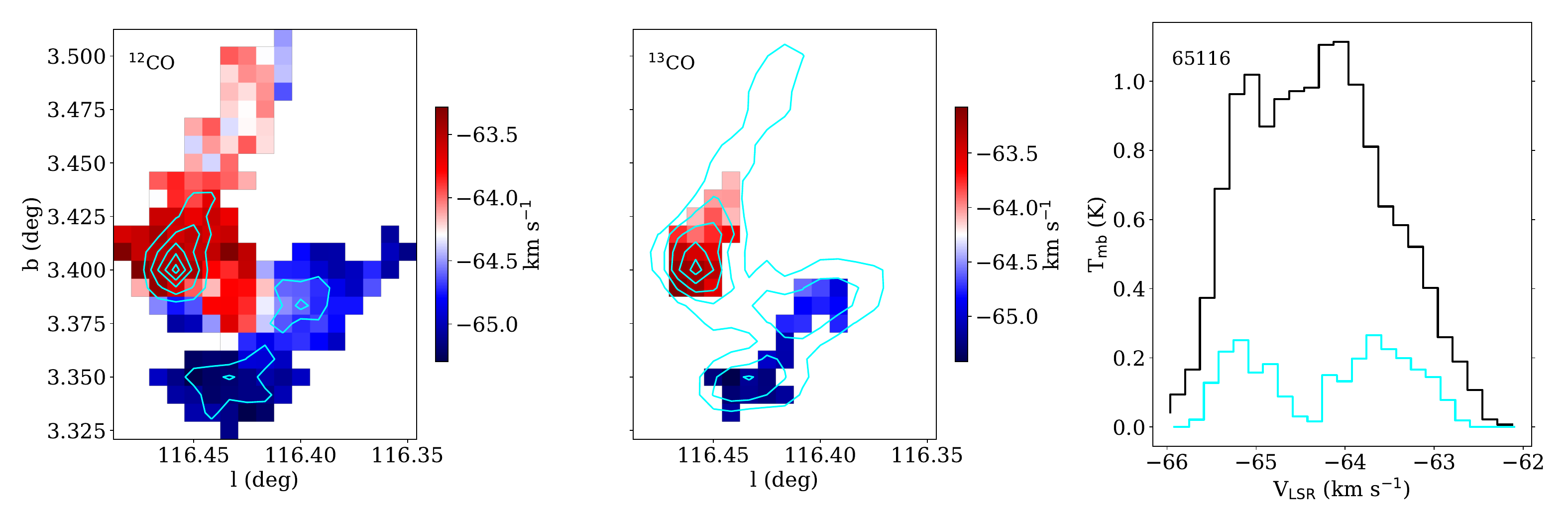}
    \plotone{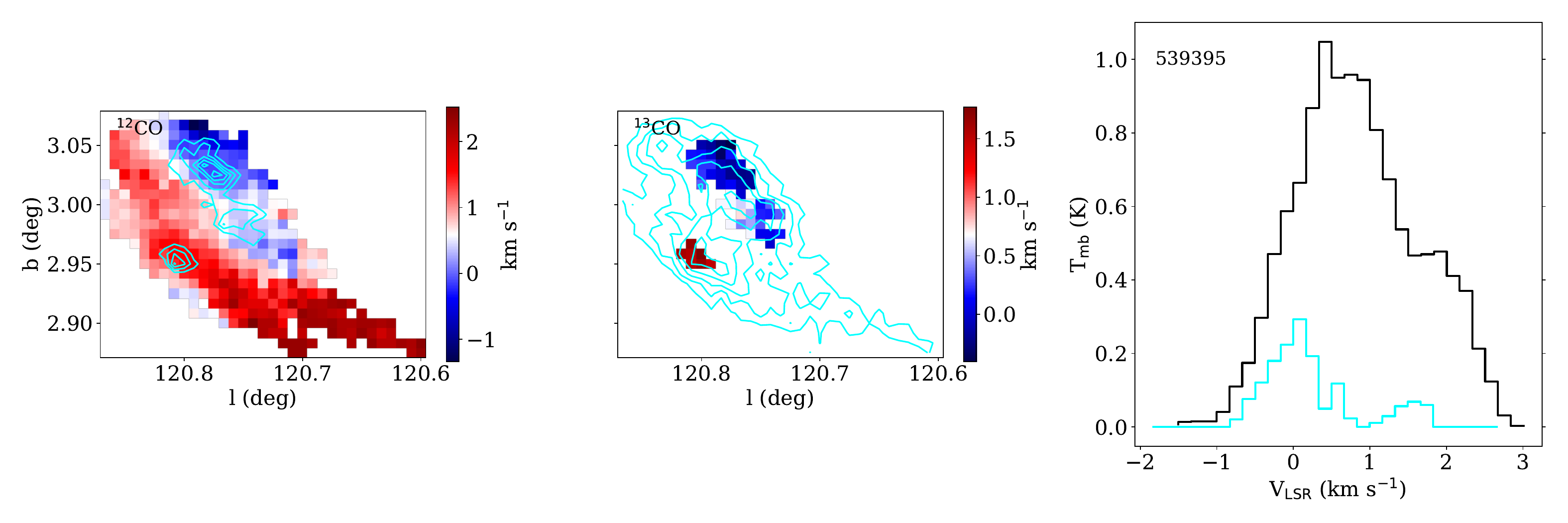}
    \plotone{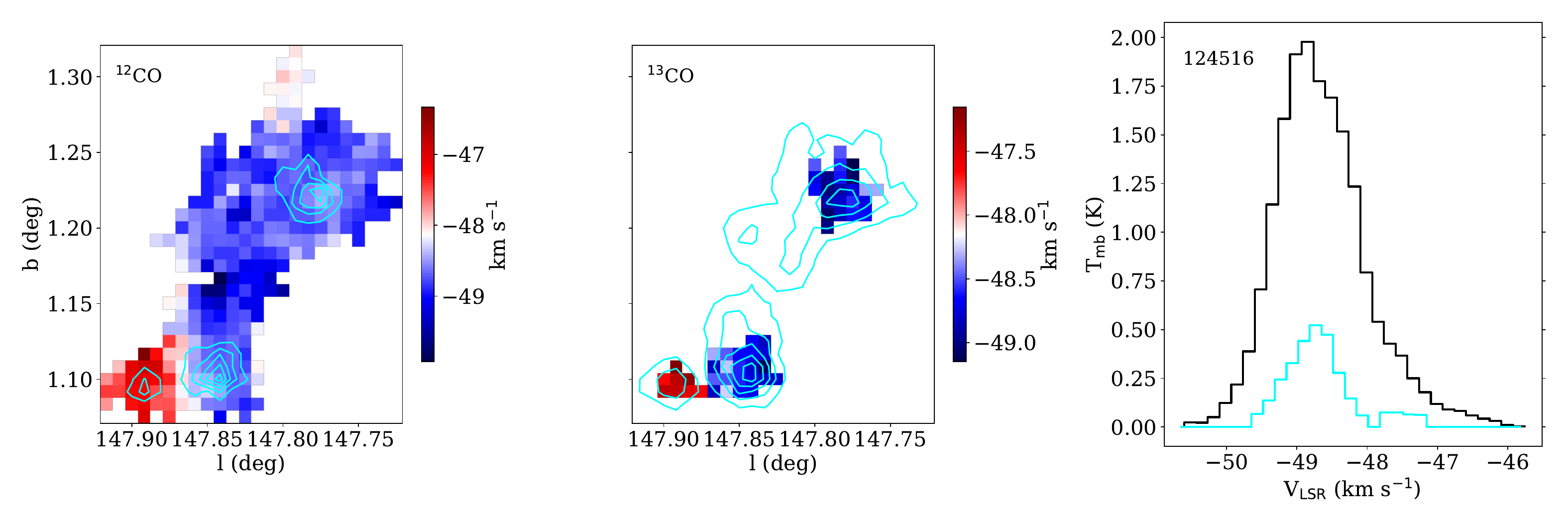}
    \plotone{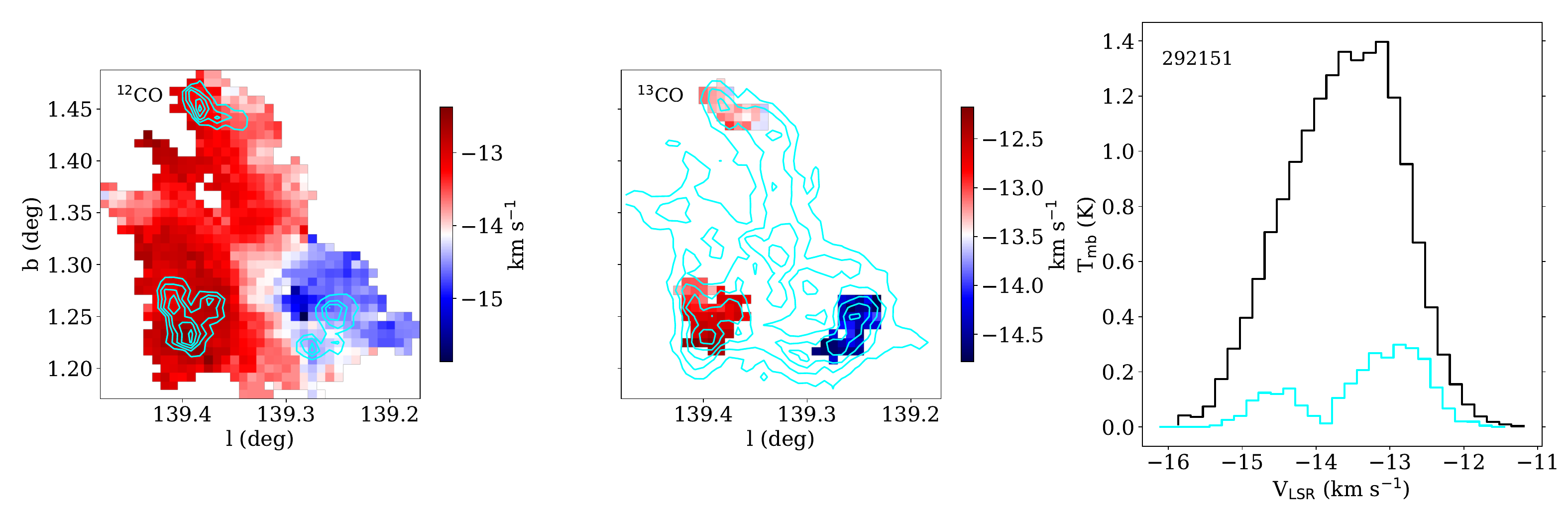}
    \caption{Same as Figure \ref{fig:f_2str}, but for the MC with triple $^{13}$CO structures, where two distinct velocity discontinuities 
    between each two of $^{13}$CO structures are determined. \label{fig:f_3str1}}
\end{figure*}

\begin{figure*}[ht]
    \plotone{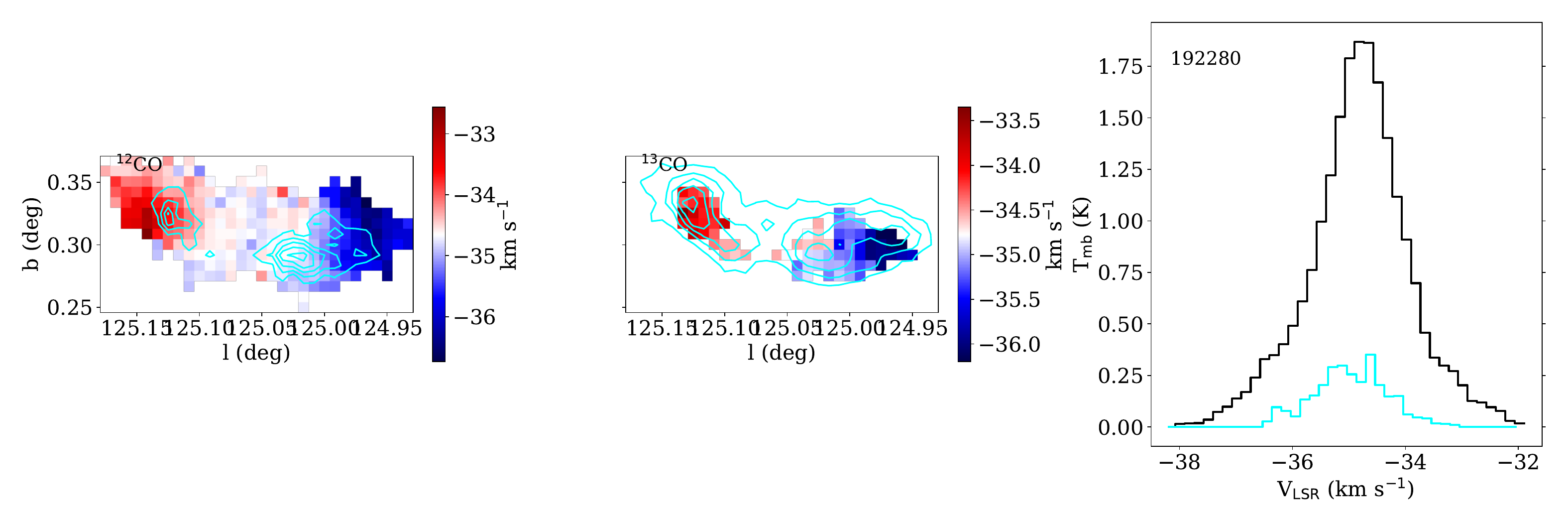}
    \plotone{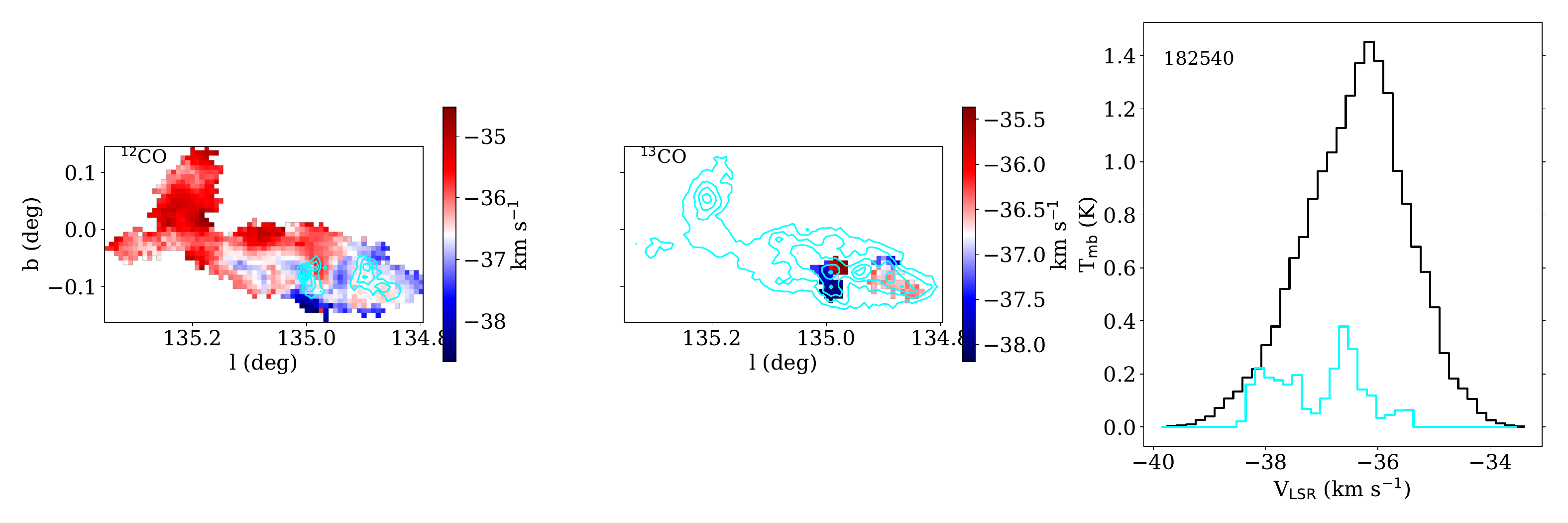}
    \plotone{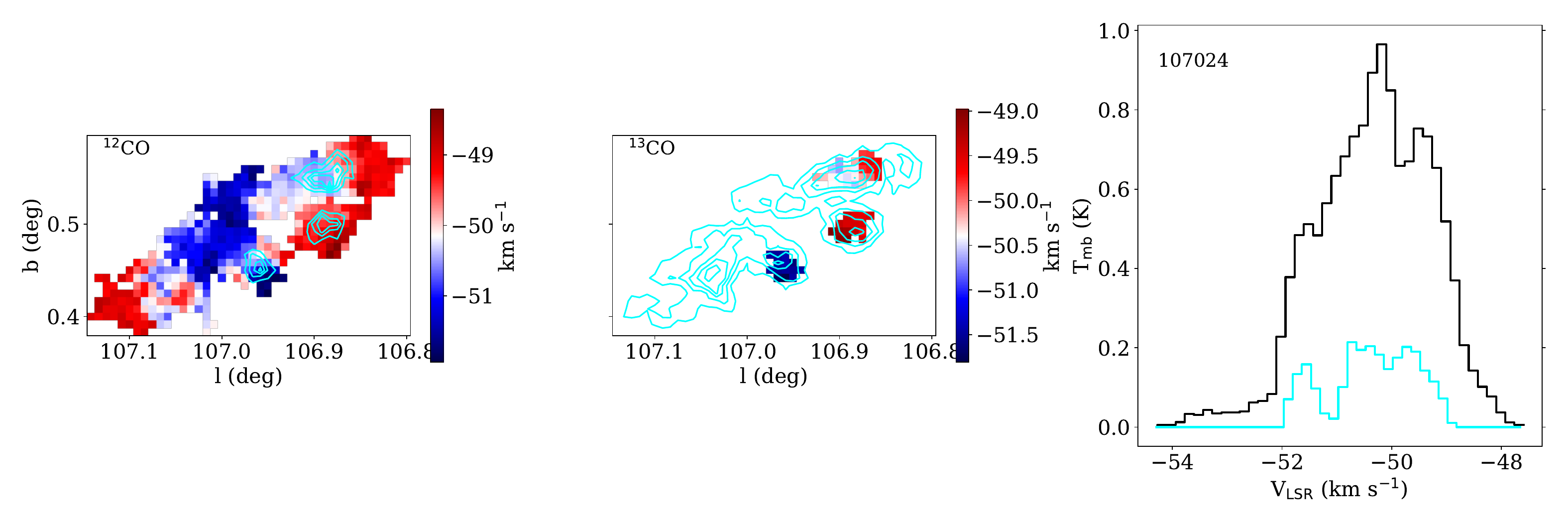}
    \plotone{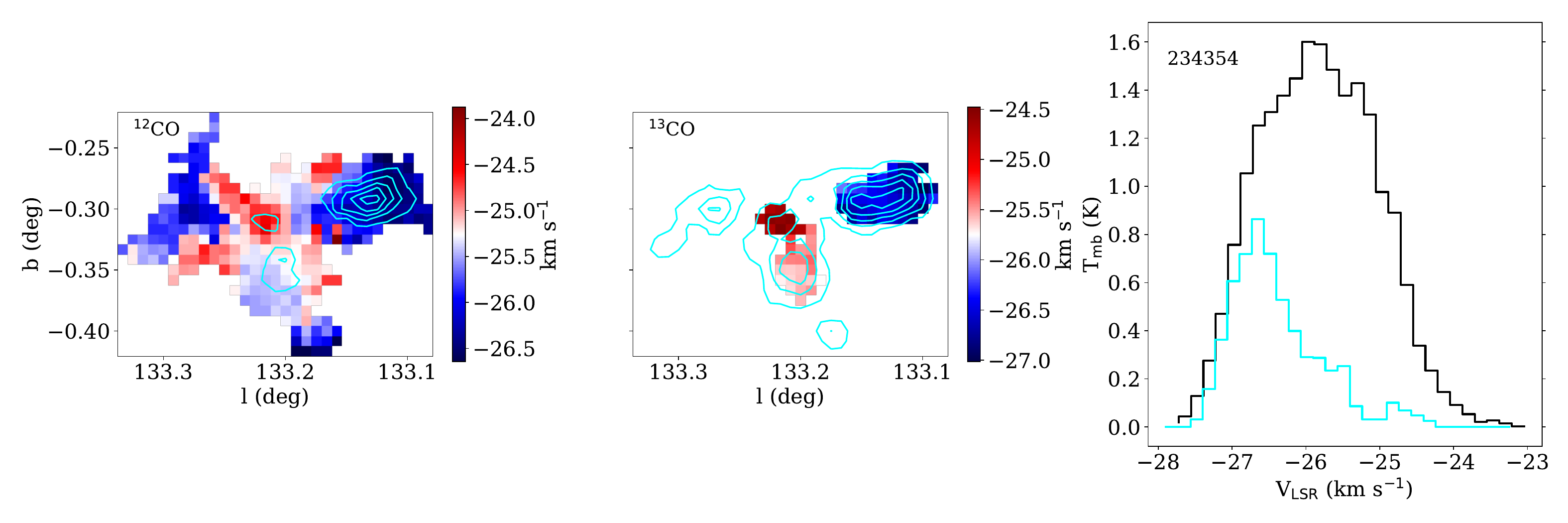}
    \caption{Same as Figure \ref{fig:f_2str}, but for the MC with triple $^{13}$CO structures, where three distinct velocities discontinuities 
    between each two of $^{13}$CO structures are determined. \label{fig:f_3str2}}
\end{figure*}

Figure \ref{fig:f_2str} shows velocity fields for MC samples with double $^{13}$CO structures, 
which display distinct velocity discontinuities. 
Figures \ref{fig:f_3str1} and \ref{fig:f_3str2} present velocity fields for MC samples with triple $^{13}$CO structures, 
which are identified to have distinct velocity discontinuities. 

\clearpage
\bibliography{velgrad_13co.bib}{}
\bibliographystyle{aasjournal}


\end{document}